\documentclass[fleqn,usenatbib]{mnras}

\usepackage{newtxtext,newtxmath}

\usepackage[T1]{fontenc}

\DeclareRobustCommand{\VAN}[3]{#2}
\let\VANthebibliography\thebibliography
\def\thebibliography{\DeclareRobustCommand{\VAN}[3]{##3}\VANthebibliography}

\usepackage{graphicx}	
\usepackage{amsmath}	

\newcommand{\fpp}[2]{\frac{\partial #1}{\partial #2}}
\newcommand{\vctr}[1]{\mbox{\boldmath $#1$}}
\newcommand{\mr}[1]{\mathrm{#1}}
\newcommand{\lara}[1]{\langle #1 \rangle}
\newcommand{\larad}[1]{\langle \langle #1 \rangle _{t} \rangle _{\mathrm{case}}}

\title[Sunspot magnetic field and energy build-up]{Impact of subsurface convective flows on the formation of sunspot magnetic field and energy build-up}

\author[T. Kaneko et al.]{
Takafumi Kaneko,$^{1,2}$\thanks{E-mail: kaneko@lmsal.com}
Hideyuki Hotta,$^{3}$
Shin Toriumi$^{4}$
and Kanya Kusano$^{5}$
\\
$^{1}$Lockheed Martin Solar and Astrophysics Laboratory, 3251 Hanover Street B/252, Palo Alto, CA 94304, USA\\
$^{2}$High Altitude Observatory, National Center for Atmospheric Research, P.O. Box 3000, Boulder, CO 80307, USA\\
$^{3}$Department of Physics, Graduate School of Science, Chiba University, 1-33 Yayoi-cho, Inage-ku, Chiba 263-8522, Japan\\
$^{4}$Institute of Space and Astronautical Science (ISAS)/Japan Aerospace Exploration Agency (JAXA),
3-1-1 Yoshinodai, Chuo-ku, Sagamihara, Kanagawa \\252-5210, Japan \\
$^{5}$Institute for Space-Earth Environmental Research, Nagoya University, Furo-cho, Chikusa-ku, Nagoya, Aichi, 464-8601, Japan
}

\date{Accepted XXX. Received YYY; in original form ZZZ}

\pubyear{2022}

\begin{document}
\label{firstpage}
\pagerange{\pageref{firstpage}--\pageref{lastpage}}
\maketitle

\begin{abstract}
  Strong solar flares occur in $\delta $-spots
  characterized by the opposite-polarity magnetic fluxes in a single penumbra.
  Sunspot formation via flux emergence from the convection zone to the photosphere
  can be strongly affected by convective turbulent flows.
  It has not yet been shown how crucial convective flows are
  for the formation of $\delta$-spots. 
  The aim of this study is to reveal the impact of convective flows
  in the convection zone on the formation and evolution of sunspot magnetic fields.
  We simulated the emergence and transport of magnetic flux tubes in the convection zone
  using radiative magnetohydrodynamics code R2D2.	
  We carried out 93 simulations by allocating the twisted flux tubes
  to different positions in the convection zone.
  As a result, both $\delta$-type and $\beta$-type magnetic distributions were
  reproduced only by the differences in the convective flows surrounding the flux tubes.
  The $\delta $-spots were formed by the collision of positive and negative magnetic fluxes
  on the photosphere. The unipolar and bipolar rotations of the $\delta$-spots
  were driven by magnetic twist and writhe, transporting magnetic helicity
  from the convection zone to the corona.
  We detected a strong correlation between the distribution of the nonpotential
  magnetic field in the photosphere and the position of the downflow plume
  in the convection zone. The correlation could be detected $20$--$30$ h
  before the flux emergence.
  The results suggest that high free energy regions
  in the photosphere can be predicted even before the magnetic flux appears
  in the photosphere by detecting the downflow profile in the convection zone.
\end{abstract}

\begin{keywords}
MHD -- sunspots -- Sun: interior -- Sun: photosphere -- Sun: magnetic fields -- Sun: flares
\end{keywords}



\section{Introduction}\label{sec:intro}

$\delta$-spots, characterized by opposite magnetic polarities in the same penumbra,
frequently produce energetic events such as solar flares and coronal mass ejections.
The relationship between the $\delta$-spots and large flares have been studied for decades
via observations of magnetic distribution and evolution
\citep{1987SoPh..113..267Z,2000ApJ...540..583S,2017ApJ...834...56T,2020Sci...369..587K}.
$\delta $-spots store a large amount of magnetic free energy
that enables intense energy release as solar flares and eruptions.
Revealing the mechanism of energy build-up in the complex magnetic distribution
of the $\delta $-spots
is key to understanding and predicting the solar activity.
Previous observations have detected the magnetic energy build-up by the emergence of new magnetic fluxes
from the convection zones \citep{1996ApJ...462..547L,2000ApJ...532..616W,2012ApJ...761...61G}
and by the converging and shearing motions between the opposite-polarity magnetic fluxes in the photosphere
\citep{1994SoPh..150..199S,2018SoPh..293..114P}.
Magnetohydrodynamic (MHD) simulations of the solar corona also demonstrated
that flux emergence and photospheric motions can build up magnetic free energy
and trigger flares and eruptions
\citep{2000ApJ...529L..49A,2000ApJ...539..954D,2007ApJ...668.1232F,2012ApJ...760...31K,2014Natur.514..465A,2014ApJ...796...44K,2016NatCo...711522J,2021ApJ...909..155K}.

The formation and evolution of $\delta $-spots have been studied
using both observational and theoretical approaches \citep{2019LRSP...16....3T}.
The properties of $\delta $-spots have been revealed via direct measurements
of the photospheric magnetic field.
However, magnetic field evolution in the convection zone
connected to the photosphere should be further studied because 
it cannot be measured directly.
MHD simulations are effective methods
to connect the evolution of magnetic field in the convection zone
to photospheric magnetic evolution.
A typical model is the twisted flux tube ascended by buoyant instability.
To date, many MHD simulations with different parameters, e.g., twist number,
buoyancy,
number of flux tubes, and background magnetic fields, have been performed
\citep{2001ApJ...554L.111F,2003ApJ...586..630M,2009ApJ...697.1529F,2015ApJ...806...79F,2015ApJ...813..112T,2017ApJ...850...39T,2021ApJ...909...72M}.
These studies have revealed that highly kinked flux tubes with two sections
of buoyant instability can reproduce the observed magnetic properties
of $\delta $-spots together with photospheric motions \citep{2014SoPh..289.3351T,2015ApJ...806...79F}.
In these simulations, the convectively unstable layer caused buoyant instability,
whereas realistic convection with radiative heat transfer was not taken into account.

Background convection affects
the transport of magnetic flux and sunspot formation.
It is difficult to demonstrate the impact of convective flows deep in the convection zone
on sunspot formation in the photosphere
because self-consistent MHD simulations covering an area from the deep convection zone to the photosphere
require large numerical resources due to the significant gaps in the typical time scale of the thermal convection and the speed of sound.
Owing to the modern numerical techniques that reduce the gap of the characteristic speeds,
sunspot formation in the photosphere can be directly reproduced
with the effect of realistic convection
in radiative MHD simulations \citep{2010ApJ...720..233C,2014ApJ...785...90R,2017ApJ...846..149C,2019ApJ...886L..21T,2020MNRAS.494.2523H,2020MNRAS.498.2925H}.
\citet{2017ApJ...846..149C} demonstrated that flux emergence is driven by convective upflows,
whereas the persistent strong magnetic field of the sunspots is formed in the subsurface downflow region.
Although their initial flux tube model was reconstructed
based on a self-consistent convective dynamo simulation
(with temporal and spatial rescaling),
the flux emergence simulation itself was carried out in a local box covering $30~\mr{Mm}$ depth;
thus, the impact of the convective flows in much deeper layers was not clear.
The flux emergence simulations covering much deeper layer have been performed \citep{2019ApJ...886L..21T,2019SciA....5.2307H,2020MNRAS.498.2925H}.
\citet{2019ApJ...886L..21T} performed a flux emergence simulation with
a deeper convection zone up to $140~\mr{Mm}$ depth, which is very close to the bottom
of the convection zone ($200~\mr{Mm}$) given the large scale heights there.
In their simulations, the flux tube was lifted upward at two sections by upflows
and dragged downward by downflows in between them.
The magnetic fluxes emerged by the upflows collided each other in the photosphere,
leading to the formation of $\delta $-spots.
In addition, the rotational shearing motion was driven between the opposite-polarity fluxes
due to the release of the magnetic twist, creating highly nonpotential fields
(arcade field hosting a flux rope along the polarity inversion line).
\citet{2020MNRAS.498.2925H} carried out a flux emergence simulation covering $200~\mr{Mm}$ depth,
reproducing the $\delta$-spots with strong horizontal fields up to $6000~\mr{G}$,
which was the maximum level in previous observations.
They performed multiple simulations by changing the spatial resolution and the approximation method of
radiation transfer (one-ray versus multi-rays). Comparison of the results showed that the amplification mechanism of the super-equipartition magnetic fields was the shearing motion
between the rotating sunspots rooted to the deep convection zone.

In this study, we examined the impact of turbulent convective flows
on the magnetic energy build-up associated with the formation of the $\delta $-spots.
We carried out a parameter survey by allocating the initial flux tube
to different positions in the identical convection field.
By analyzing the large number of simulated cases,
we revealed the distribution trend of the nonpotential magnetic field in the photosphere
and its relationship with the flow distribution in the convection zone.
The remainder of the paper is structured as follows.
Section 2 describes the setting of the numerical simulation.
Section 3 shows the results of the calculations and statistical analysis.
In Section 4, we summarize and discuss the results.

\section{Numerical Settings} \label{sec:set}
We solved the three-dimensional MHD equations including the radiative transfer equation.

\begin{equation}
  \fpp{\rho _{1}}{t}=-\frac{1}{\xi ^{2}}\nabla \cdot \left( \rho \vctr{v} \right),
\end{equation}
\begin{equation}
  \fpp{(\rho \vctr{v})}{t}=-\nabla \cdot \left( \rho \vctr{v}\vctr{v} \right)-\nabla p_{1}
  +\rho _{1}\vctr{g}+\frac{1}{4\pi }\left( \nabla \times \vctr{B} \right)\times \vctr{B},
\end{equation}
\begin{equation}
  \fpp{\vctr{B}}{t}=\nabla \times \left( \vctr{v}\times \vctr{B} \right),
\end{equation}
\begin{equation}
 \rho T \fpp{s_{1}}{t}=\rho T \left( \vctr{v}\cdot \nabla \right)s+Q,
\end{equation}
\begin{equation}
  p_{1}=p_{1}\left( \rho,s \right), \label{eq:eos}
\end{equation}
\begin{equation}
  \rho = \rho _{0}+\rho _{1},
\end{equation}
\begin{equation}
  p = p_{0}+p_{1},
\end{equation}
\begin{equation}
  s = s_{0}+s_{1},
\end{equation}
where $\rho $, $\vctr{v} $, $\vctr{B} $, $p$, $T$, $s$, $\vctr{g}$, $Q$, and $\xi $
are the density, fluid velocity, magnetic field, gas pressure, temperature, entropy, gravitational acceleration in the vertical direction, radiative heating,
and the factor of the reduced speed of sound technique (RSST), respectively. The subscript $0$ denotes the variables of the stationary stratification in $z$-direction,
and the subscript $1$ denotes the perturbation.
The background stratification was calculated using the hydrostatic equation based on Model S \citep{1996Sci...272.1286C}. In Eq. (\ref{eq:eos}),
we used the equation of state considering the partial ionization effect with the OPAL repository \citep{1996ApJ...456..902R}.
See \citet{2020MNRAS.494.2523H} for the details of the calculation procedure.

The calculations were carried out using the radiative MHD code R2D2
\footnote{Radiation and RSST for Deep Dynamics} \citep{2019SciA....5.2307H}.
The MHD equations are solved using the four-step Runge Kutta method for time integration
\citep{2005A&A...429..335V} and fourth-order spatial derivative
\citep{2020MNRAS.494.2523H} with the slope-limited artificial diffusion \citep{2014ApJ...789..132R}.
The radiative transfer equation was solved by adopting the gray approximation and the Rosseland mean opacity.
We only solved two rays for the radiative transfer equation: only upward and downward radiative energy transfers were considered. The horizontal radiative energy transfer should be included to reproduce the realistic solar convection by diffusing the horizontal small scale structures. In our simulations with a low spatial resolution, on the other hand, the realistic convection was eventually mimicked via the numerical diffusion. The high resolution simulations \citep[e.g.,][]{2020MNRAS.498.2925H} should include the horizontal radiative energy transfer because the numerical diffusion is reduced.
The RSST relaxed the Courant–Friedrichs–Lewy (CFL) condition dependent on the high sound speed
in the photosphere \citep{2012A&A...539A..30H,2015ApJ...798...51H,2019A&A...622A.157I}.
We set the RSST factor as follows:
\begin{equation}
  \xi \left(z \right)=\max \left( 1, \xi _{0}\left[ \frac{\rho _{0}(z)}{\rho _{b}}\right]^{1/3} \frac{c_{s}(z)}{c_{b}} \right),
\end{equation}
where $\xi _{0}=160$ was adopted.
$\rho _{b}=0.2~\mathrm{g~cm^{-3}}$ and $c_{b}=2.2\times 10^{7}~\mathrm{cm~s^{-1}}$ were the density and sound speed around the base of the convection zone, respectively.
$c_{s}=\sqrt{\left(\partial p/\partial \rho \right)_{s}}$ is the local adiabatic speed of sound.
We also limited the Alfv\'en speed up to $40~\mr{km/s}$ to deal with the low-$\beta $ region
above the photosphere \citep{2009ApJ...691..640R}.

The simulation domain is a rectangular box with the size of
$0<x<98.3~\mr{Mm}$, $0<y<98.3~\mr{Mm}$, and $-201~\mr{Mm}<z<676 ~\mr{km}$ in Cartesian coordinates.
The $x$--$y$ plane represents the horizontal plane parallel to the solar surface,
and the $z$-direction represents the height. $z=0$ corresponds to $R=R_{\mr{sun}}$
, where $R$ and $R_{\mr{sun}}$ represent the distance from the center of the Sun and the solar radius, respectively.
To inhibit the convection cell lager than the solar supergranulation scale,
  we limit the horizontal size of the simulation domain
  within $100~\mr{Mm}$ (see also \citet{2019ApJ...886L..21T,2019SciA....5.2307H}).
The simulation domain covers the area from the deep convection zone to the lower chromosphere.
In the vertical direction,
uniform grid spacing was applied from the top boundary to $z=-4~\mr{Mm}$ with a grid size of $\Delta z=48~\mr{km}$.
The grid size linearly increased from $z=-4~\mr{Mm}$ to the bottom boundary up to $\Delta z=903~\mr{km}$.
Uniform grid spacing was applied in the horizontal direction
with a grid size of $\Delta x=\Delta y=128~\mr{km}$.
The total grid number was $768 \times 768 \times 512$.
The periodic boundary condition was applied to the horizontal direction.
The magnetic field in the bottom boundary was horizontal. 
The magnetic field in the top boundary was connected to the potential field.

\begin{figure}
  \begin{center}
    \includegraphics[width=\columnwidth]{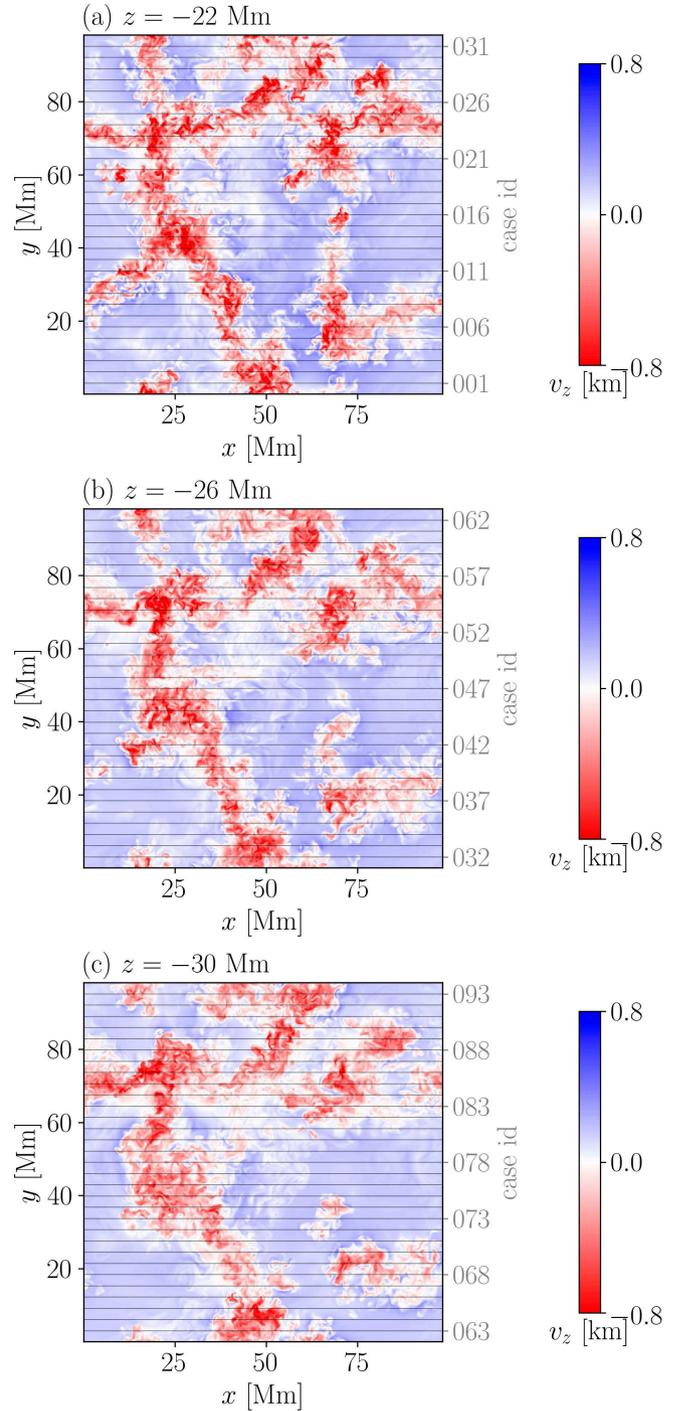}
    \caption{Initial positions of the flux tubes in the background vertical velocity.
      The gray lines represent the axial position of the flux tube in each case.
      The color represents vertical velocity $v_{z}$.}
    \label{fig:tube_pos}
  \end{center}
\end{figure}

First, we calculated the solar convection without a magnetic field
until statistical equilibrium was achieved. We called this state as the initial state.
Then, we introduced a magnetic flux tube into the convection zone.
The magnetic field of the flux tube was calculated as a horizontal force-free field as follows:
\begin{equation}
  B_{x}=B_{\mr{tb}}J_{0}(\alpha r),~B_{\phi }=B_{\mr{tb}}J_{1}(\alpha r),
\end{equation}
where $B_{x}$ and $B_{\phi }$ are the toroidal and poloidal components, respectively.
$r$ is the radial distance from the axis, $B_{\mr{tb}}=10^{4}~\mr{G}$ is the field strength
of the axial magnetic field, $J_{0}$ and $J_{1}$ are the Bessel functions,
$\alpha = a_{0}/R_{\mr{tb}}$, $a_{0}=2.404825$, $R_{\mr{tb}}=8.5~\mr{Mm}$.
The flux tube had right-handed twist without writhe in the initaial state.
We carried out 93 simulations by allocating the initial flux tubes to 93 different positions.
The initial positions of the flux tubes are summarized in Table \ref{table:case}.
Figure \ref{fig:tube_pos} shows the initial positions of the flux tube axes
against the background vertical velocity.
The horizontal position $y_{\mr{tb}}$ was uniformly changed in the range of
$24\Delta y$ --$744\Delta y$ ($3.072~\mr{Mm}$--$95.232~\mr{Mm}$)
with an interval of $24\Delta y$ ($3.072~\mr{Mm}$),
corresponding to 31 different cases at the same depth.
Note that the interval of $3~\mathrm{Mm}$ was selected
because the downflows had the small structures at this spatial scale
as shown in Fig. \ref{fig:tube_pos}.
The depth $z_{\mr{tb}}$ was set to $=-22~\mr{Mm}$ (cases 001--031), $-26~\mr{Mm}$ (cases 032--062),
and $-30~\mr{Mm}$ (cases 063--093).
As shown in Fig. \ref{fig:tube_pos}, the flow profiles at these depths were similar
to each other. We show that the magnetic evolution from the different depths
also resulted in a variety of the sunspot magnetic fields.

\begin{table}
	\centering
	\caption{Initial positions of the flux tubes}
	\label{table:case}
	\begin{tabular}{ccc} 
		\hline
		 case & $z_{\mr{tb}}$ & $y_{\mr{tb}}$ \\
		\hline
        $001$--$031$ & $-22~\mr{Mm}$ & $3.072~\mr{Mm}$--$95.232~\mr{Mm}$ \\
        $032$--$062$ & $-26~\mr{Mm}$ & $3.072~\mr{Mm}$--$95.232~\mr{Mm}$ \\
        $063$--$093$ & $-30~\mr{Mm}$ & $3.072~\mr{Mm}$--$95.232~\mr{Mm}$ \\
		\hline
	\end{tabular}
\end{table}

\section{Results} \label{sec:res}
\subsection{Typical case}
\begin{figure}
  \begin{center}
    \includegraphics[width=\columnwidth]{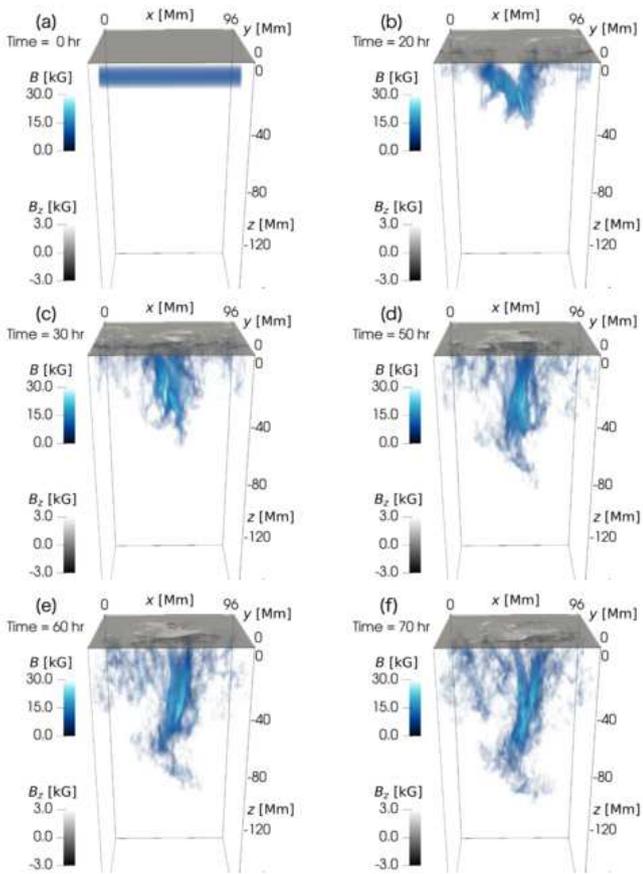}
    \caption{Temporal evolution of magnetic field in case 001.
      The grayscale represents the vertical magnetic field $B_{z}$ at the $z=0$ plane.
      Blue represents the field strength $|\vctr{B}|$.
      Only the region where $|\vctr{B}|>3~\mr{kG}$ is rendered.
      For better visualization, the coordinate is horizontally translated
      as the initial flux tube position corresponds to the center in the $y$-direction.}
    \label{fig:mag_t}
  \end{center}
\end{figure}

\begin{figure}
  \begin{center}
    \includegraphics[width=\columnwidth]{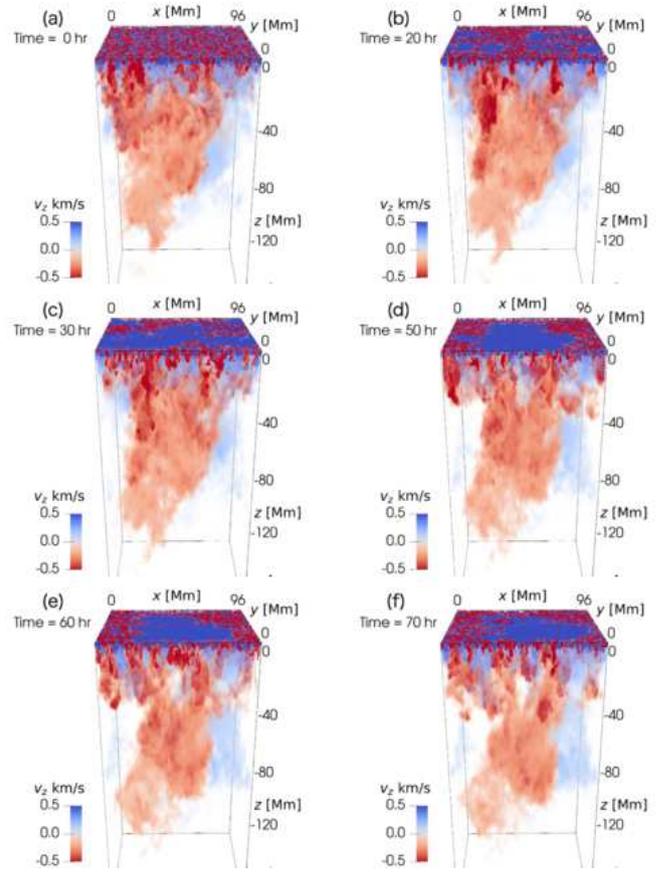}
    \caption{Temporal evolution of vertical flows $v_{z}$ in case 001.
      Blue and red represent upflow and downflow, respectively.
      Only the regions where $|v_{z}|>0.1~\mr{km/s}$ are rendered.
      The coordinate is the same as in Fig. \ref{fig:mag_t}.}
    \label{fig:vz_t}
  \end{center}
\end{figure}

Figure \ref{fig:mag_t} represents the temporal evolution of the magnetic field in case 001
where $\delta $-type magnetic distribution with a large spot area was reproduced.
In this case, the opposite-polarity magnetic fluxes in the photosphere collided with each other,
forming the $\delta $-type magnetic distribution.
Figure \ref{fig:vz_t} represents the temporal evolution of the vertical flows in case 001.
The convective velocity field in our simulation included a persistent large-scale
downflow plume extending to the deep layer.
The downflow plume in the subsurface layer was also reproduced
in the previous simulations \citep{1989ApJ...342L..95S}.
Dragged by the downflow plume,
the flux tube was deformed into the concave-up structure (Fig. \ref{fig:mag_t}(b)),
and gradually sank to the deep convection zone.
Meanwhile, the opposite-polarity magnetic fluxes in the photosphere
approached each other and finally collided.

\begin{figure}
  \begin{center}
    \includegraphics[width=\columnwidth]{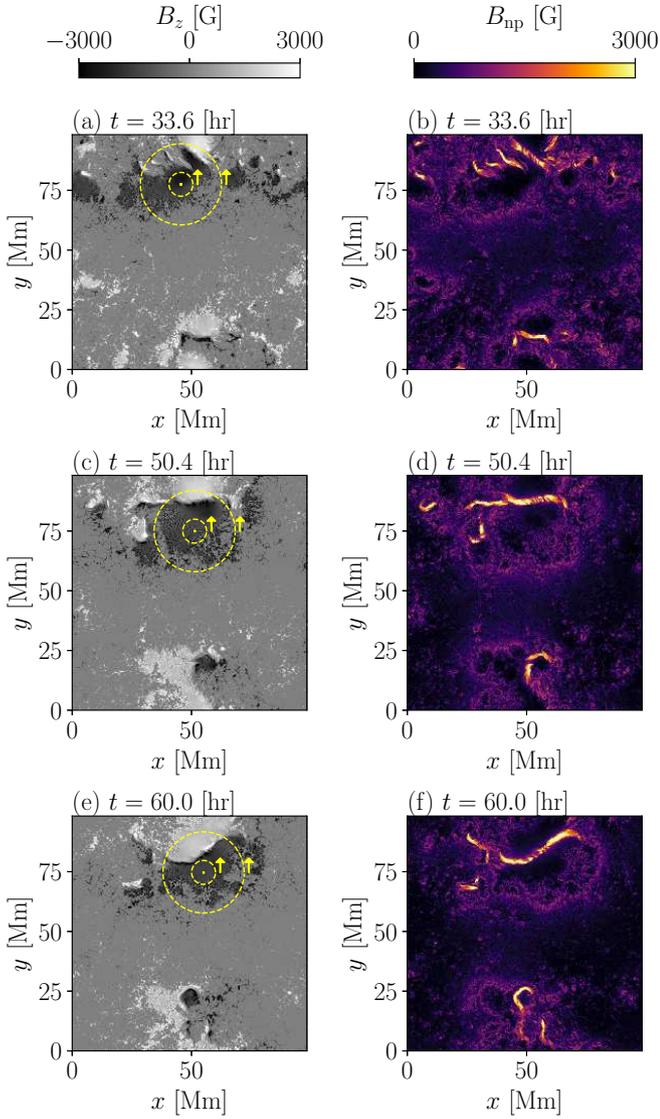}
    \caption{Panels (a)--(f) shows temporal evolution of $B_{z}$ and $B_{\mr{np}}$
        at the $\tau=1$ surface in case 001.
        The yellow dashed cicles are the slits used for Fig. \ref{fig:tslice}.}
    \label{fig:bnp001}
  \end{center}
\end{figure}

\begin{figure}
  \begin{center}
    \includegraphics[width=\columnwidth]{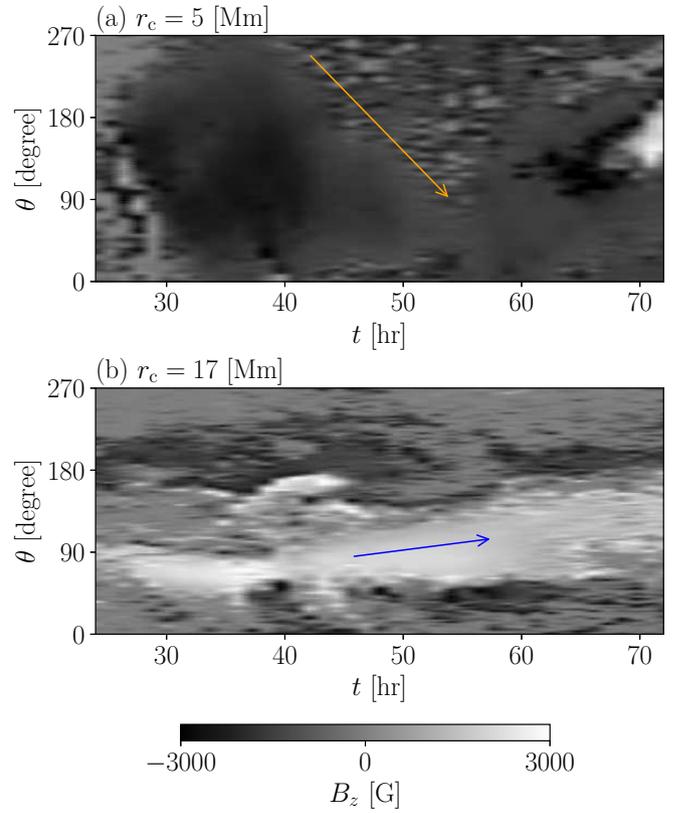}
    \caption{Time slice of the $B_{z}$ map along the circular slits
        denoted by yellow dashed circles in Fig. \ref{fig:bnp001} (a), (c) and (e).
        Panel (a) and (b) show the results with the slit radii
        $r_{\mathrm{c}}=5~\mathrm{Mm}$ and $r_{\mathrm{c}}=17~\mathrm{Mm}$, respectively.}
    \label{fig:tslice}
  \end{center}
\end{figure}

Figures \ref{fig:bnp001} (a), (c), and (e) show the temporal evolution
of the vertical magnetic field at the $\tau =1$ surface.
Figure \ref{fig:tslice} (a) and (b) show the time slice of $B_{z}$ map along
the circular slits with the different radii
denoted by the yellow dashed circles in panels (a), (c), and (e).
The center of the circular slits corresponds to the flux weighted centroid
of the negative flux in each snapshot.
Figure \ref{fig:tslice} (a) shows
the clockwise rotational motion inside the negative flux (unipolar rotation)
from the emergence time to $t=55~\mathrm{h}$.
The clockwise rotation at the photosphere reduced the positive magnetic helicity
of the flux tube in the convection zone, and increased the positive helicity
over the photosphere, transporting the positive helicity from the convection zone to the corona.
This motion is consistent with the previous numerical experiments without
convection \citep{2009ApJ...697.1529F,2015A&A...582A..76S}
and with convection \citep{2019ApJ...886L..21T}.

\begin{figure}
  \begin{center}
    \includegraphics[width=\columnwidth]{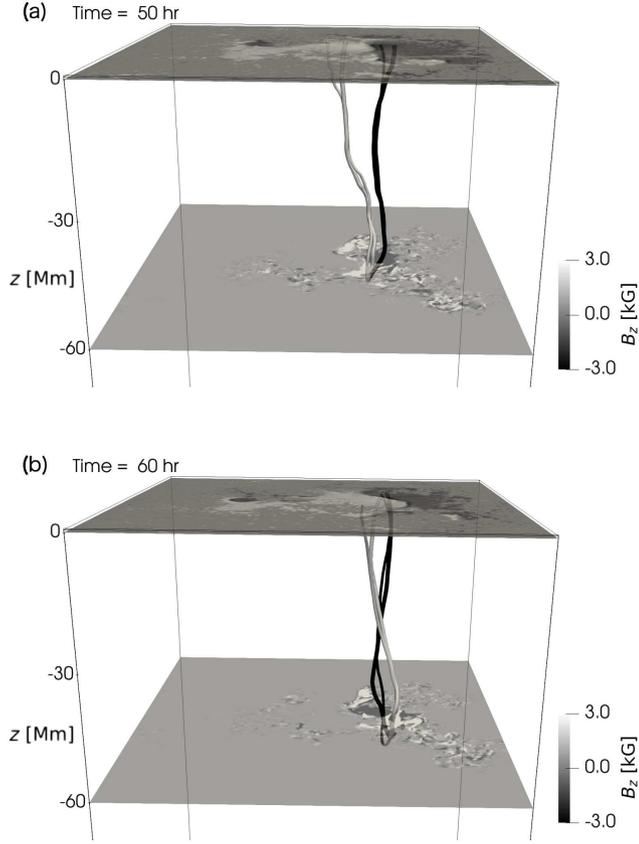}
    \caption{Evolution of magnetic field writhe in the convection zone.}
    \label{fig:writhe}
  \end{center}
\end{figure}

Figure \ref{fig:tslice} (b) represent the anti-clockwise rotation of the positive flux against
the flux centroid of the negative flux (bipolar rotation).
The driver of the anti-clockwise rotation was the conversion of the twist to the writhe.
Figure \ref{fig:writhe} shows the temporal evolution of the magnetic field lines
in the convection zone during the anti-clockwise rotation.
The positive writhe was created between the magnetic field lines
connecting to the positive and negative fluxes at the photosphere.
The helicity of the flux tube is described as the contributions from twist and writhe as follows,
\begin{equation}
  \mathcal{H}=\left( \mathcal{T}_{w}+\mathcal{W}_{r}\right)\Phi ^{2},
\end{equation}
where $\mathcal{H}$, $\mathcal{T}_{w}$, and $\mathcal{W}_{r}$ represent helicity, twist, and writhe, respectively \citep{2006JPhA...39.8321B}.
The initial straight flux tube had the positive helicity in the form of the right-handed twist.
The right-handed twist is converted to the positive writhe as the positive and negative ends
of the bending flux tube approached each other. The anti-clockwise motion between the positive
and negative fluxes at the photosphere was driven by the increasing of the positive writhe.

As a proxy of flare productivity,
we used nonpotential magnetic field defined as follows:
\begin{equation}
  B_{\mr{np}}=|\vctr{B}-\vctr{B}_{\mr{pot}}|,
\end{equation}
where $\vctr{B}_{\mr{pot}}$ represents the potential field computed
from the vertical component of magnetic field at the $\tau =1$ surface.
Note that, to mimic the analysis of the observational data,
the height difference in the $\tau =1 $ surface (within $300~\mathrm{km}$) was not considered
in the computation of $\vctr{B}_{\mr{pot}}$. 
Since the vertical components of $\vctr{B}$ and $\vctr{B}_{\mr{pot}} $ are the same,
$B_{\mr{np}}$ represents the difference in the horizontal components
between $\vctr{B}$ and $\vctr{B}_{\mr{pot}}$.
$B_{\mr{np}}$ is also associated with the magnetic free energy
responsible for the maximum energy released by flares.
Figures \ref{fig:bnp001} (b), (d), and (f) show the spatial distribution
and temporal evolution of $B_{\mr{np}}$ at the $\tau =1$ surface.
Strong nonpotential fields were localized around the polarity inversion lines.
This feature was common to all the cases of parameter survey in this study,
and to the actual active regions \citep{2020Sci...369..587K}.

\subsection{Parameter Survey}
\begin{figure}
  \begin{center}
    \includegraphics[width=\columnwidth]{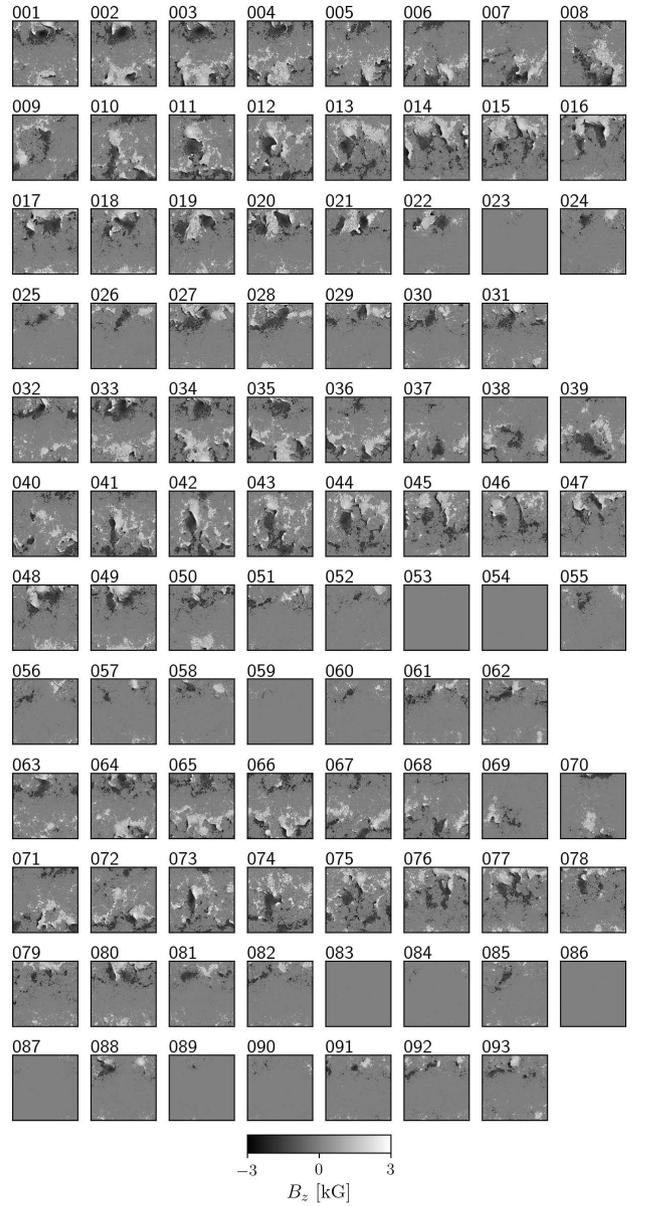}
    \caption{$B_{z}$ at the $\tau =1$ surface of all cases
      at the time when the amount of magnetic flux was maximum in each case.}
    \label{fig:all_case}
  \end{center}
\end{figure}

We carried out 93 simulations with different initial flux tube positions.
Figure \ref{fig:all_case} displays the distribution of the magnetic flux at the $\tau =1$ surface
for all 93 cases.
Each panel shows the snapshot when the amount of flux was maximum in each case.
We showed that various magnetic distributions are created only
by the difference in the convective flows surrounding the flux tubes.
We can find clear differences even between the cases
in which the horizontal positions of the initial flux tubes were the same
but the depths were different.
In many cases, opposite-polarity fluxes collided in the photosphere,
forming $\delta$-type magnetic distribution.
Some cases reproduced $\beta$-type magnetic distribution.
Several cases resulted in no successful emergence to the photosphere.
In cases where the most of the initial flux tube was occupied by a downflow,
the flux tube sank to the deep layer and never emerged to the photosphere.

\begin{figure}
  \begin{center}
    \includegraphics[width=\columnwidth]{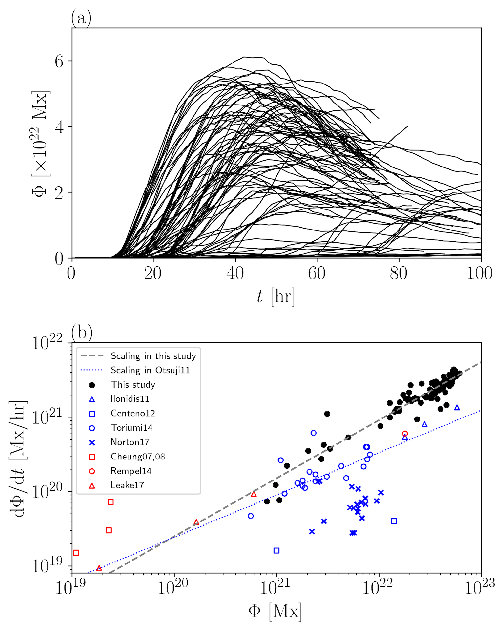}
    \caption{Panel (a) shows the temporal evolution of the unsigned magnetic flux $\Phi $ in each case.
      Panel (b) is the scatter plot of $\max(\Phi )$ versus $\max(d\Phi /dt)$.
      The plot shows the results of each case in this study (black circles),
      previous observations \citep[red marks,][]{2011Sci...333..993I,2012ApJ...759...72C,2014ApJ...794...19T,2017ApJ...842....3N}, and previous simulations \citep[blue marks,][]{2007A&A...467..703C,2008ApJ...687.1373C,2014ApJ...785...90R,2017ApJ...838..113L}.
      The gray dashed line represents the scaling fitted to the results in this study.
      The blue dotted line represents the scaling in a previous observational study
      that analyzed over 100 samples \citep{2011PASJ...63.1047O}.}
    \label{fig:flux}
  \end{center}
\end{figure}

Figure \ref{fig:flux} (a) shows the temporal evolution of the unsigned magnetic flux
at the $\tau =1$ surface defined as follows:
\begin{equation}
  \Phi = \int _{\tau=1} \left| B_{z} \right| dxdy.
\end{equation}
Here we regarded the $\tau=1$ surface as the horizontal plane
because of the small height difference.
In most cases, temporal profiles can be divided into emergence and decay phases, which are
consistent with the previous observational and numerical studies \citep{2011Sci...333..993I,2011PASJ...63.1047O,2012ApJ...759...72C,2014ApJ...794...19T,2017ApJ...842....3N,2007A&A...467..703C,2008ApJ...687.1373C,2014ApJ...785...90R}.
We only focused on the emergence phase here because the decay phase was not covered in several cases
during the calculation time.
The black circles in Fig. \ref{fig:flux} (b) represent
$\max (\Phi )$ vs $\max (d\Phi /dt )$ in our simulations.
The results revealed the scaling relationship of $d\Phi /dt \propto \Phi ^{0.78}$.
Figure \ref{fig:flux} (b) also shows the results
of the previous observational and numerical studies
(the references are summarized in \citet{2017ApJ...842....3N}).
The magnetic fluxes in our study covered relatively larger values than the results of the previous studies.
The flux emergence rates in our results were several times larger than the observational values
in the cases of the larger flux amount. 
Thus, the scaling exponent $0.78$ in our results was larger than the observational value $0.57$
reported by \citet{2011PASJ...63.1047O}.

\begin{figure}
  \begin{center}
    \includegraphics[width=\columnwidth]{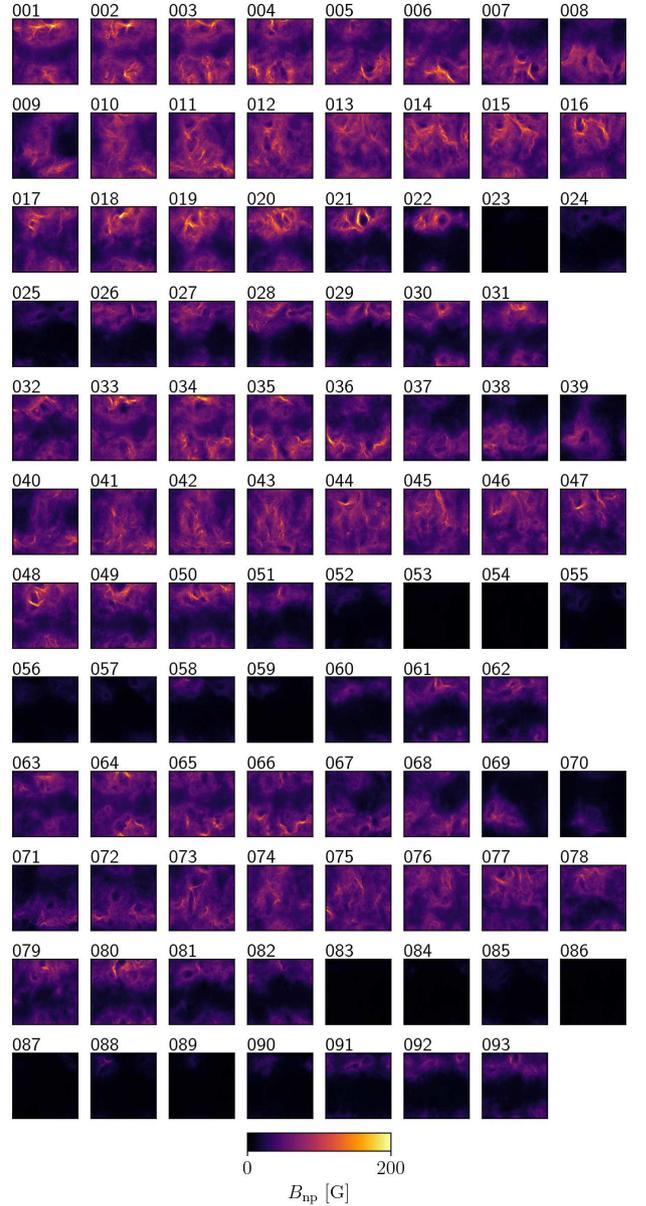}
    \caption{Temporally averaged nonpotential field $\lara{B_{\mr{np}}}_{t}$ at the $\tau =1$ surface of all cases.}
    \label{fig:bnps_each}
  \end{center}
\end{figure}

\begin{figure}
  \begin{center}
    \includegraphics[width=\columnwidth]{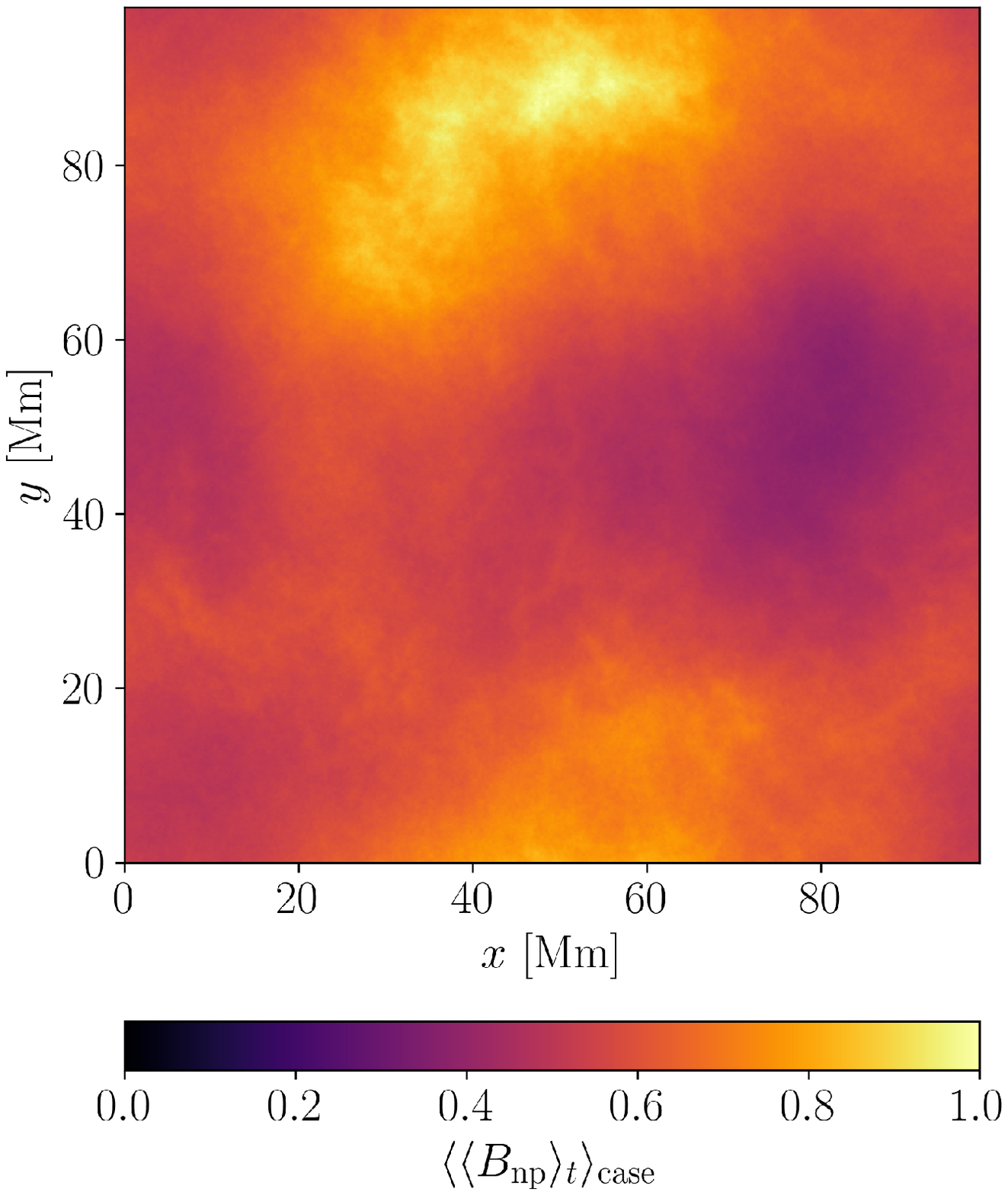}
    \caption{Distribution of $\larad{B_{\mr{np}}}$ (see Eqs. (\ref{eq:bnpt})--(\ref{eq:bnpf}).}
    \label{fig:bnps}
  \end{center}
\end{figure}

We investigated the distribution trend of the magnetic nonpotential field in the photosphere.
First, we derived the temporally averaged nonpotential field in each case:
\begin{equation}
  \lara{B_{\mr{np}}}_{t}=\frac{1}{t_{\mr{c}}}\int_{0}^{t_{\mr{c}}} B_{\mr{np}}(t,y,z)dt,
  \label{eq:bnpt}
\end{equation}
where $t$ and $t_{\mr{c}}$ represent the time and duration of the simulation in each case, respectively.
Note that $t_{\mr{c}}$ is different in each case.
Figure \ref{fig:bnps_each} displays $\lara{B_{\mr{np}}}_{t}$ in each case.
To evaluate the trend of the $B_{\mr{np}}$ distribution,
we summed $\lara{B_{\mr{np}}}_{t}$ of all cases and normalized the sum:
\begin{equation}
  \larad{B_{\mr{np}}} = F/F_{\mr{max}},
  \label{eq:bnptc}
\end{equation}
where
\begin{equation}
  F=\sum _{\mr{case}} \lara{B_{\mr{np}}}_{t},
  \label{eq:bnpf}
\end{equation}
and $F_{\mr{max}}$ is the maximum value of $F$.
Figure \ref{fig:bnps} shows $\larad{B_{\mr{np}}}$.
The regions where $\larad{B_{\mr{np}}}$ is close to unity have a higher possibility
that a strong nonpotential field is created.

\begin{figure}
  \begin{center}
    \includegraphics[width=\columnwidth]{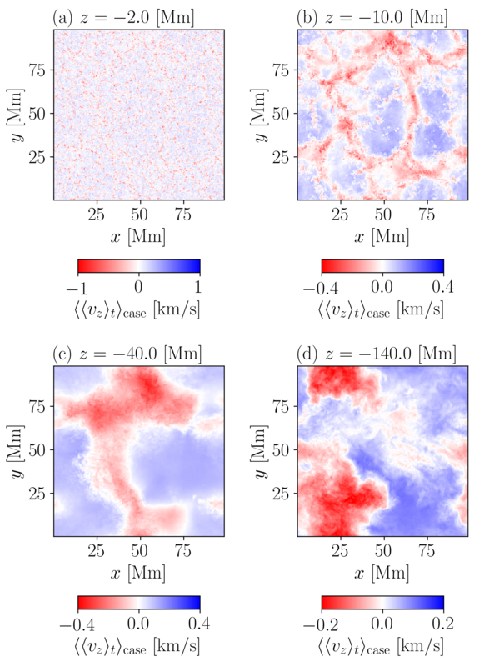}
    \caption{Mean vertical velocity $\larad{v_{z}}$ at different heights
      (see Eqs. (\ref{eq:vt})--(\ref{eq:vtc})).}
      \label{fig:vxs}
  \end{center}
\end{figure}

\begin{figure}
  \begin{center}
    \includegraphics[width=\columnwidth]{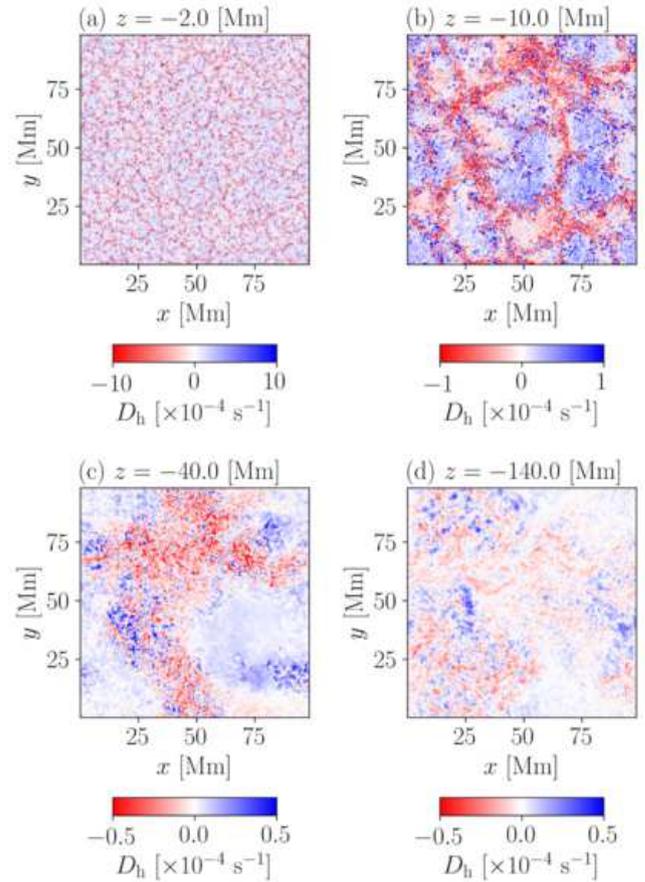}
    \caption{Horizontal divergence of mean velocity field $D_{\mr{h}}$ at different heights
      (see Eq. (\ref{eq:dh})).}
      \label{fig:div}
  \end{center}
\end{figure}

\begin{figure}
  \begin{center}
    \includegraphics[width=\columnwidth]{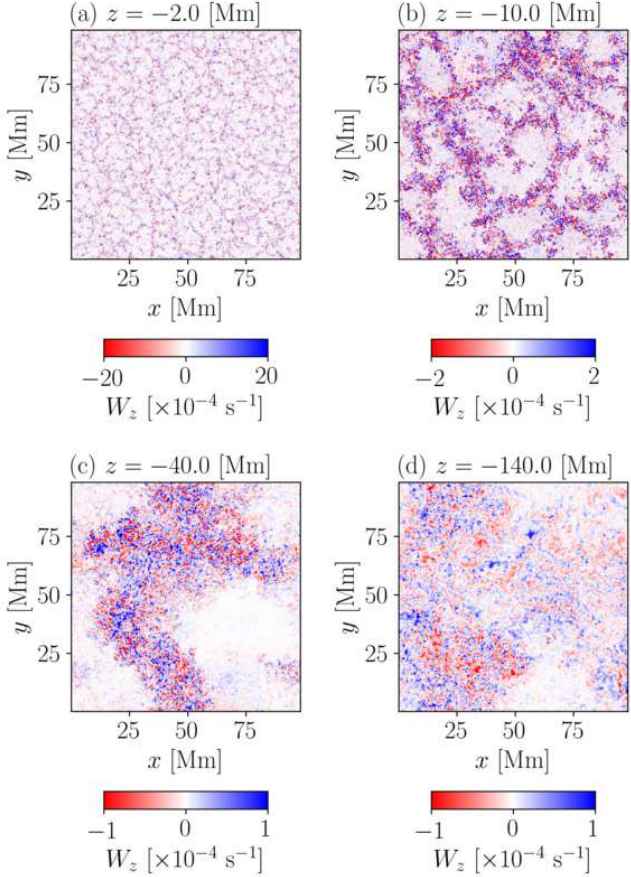}
    \caption{Vertical vorticity of mean velocity field $W_{z}$ at different heights
      (see Eq. (\ref{eq:wz})).}
      \label{fig:rot}
  \end{center}
\end{figure}

\begin{figure}
  \begin{center}
    \includegraphics[width=\columnwidth]{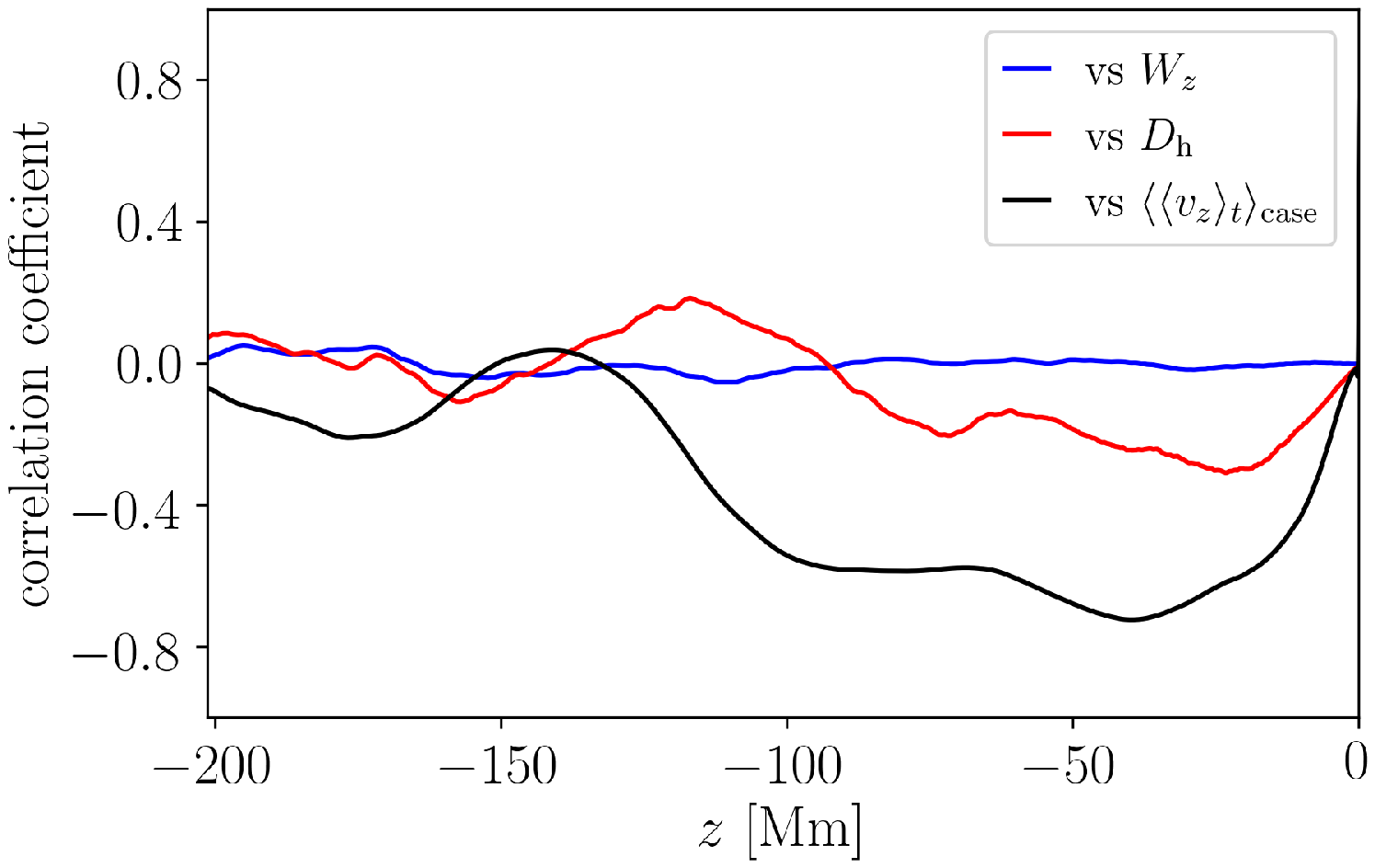}
    \caption{Correlation coefficients of
      $\larad{v_{z}}$ vs $\larad{B_{\mr{np}}}$ (black line),
      $D_{\mr{h}}$ vs $\larad{B_{\mr{np}}}$ (red line),
      $W_{z}$ vs $\larad{B_{\mr{np}}}$ (blue line) as a function of height.}
      \label{fig:cor}
  \end{center}
\end{figure}

To investigate the cause of the $\larad{B_{\mr{np}}}$ distribution,
we compared it with the mean velocity field in the convection zone.
The mean velocity field $\larad{v_{q}}$ was computed as follows:
\begin{equation}
  \larad{v_{q}}
  = \frac{1}{N}\sum _{\mr{case}} \lara{v_{q}}_{t},
  \label{eq:vt}
\end{equation}
\begin{equation}
  \lara{v_{q}}_{t}=\frac{1}{t_{\mr{c}}}\int_{0}^{t_{\mr{c}}} v_{q}(t,x,y,z)dt,
  \label{eq:vtc}
\end{equation}
where $q$ denotes the component $x$, $y$, or $z$, and $N=93$ is the number of the cases.
Figure \ref{fig:vxs} shows $\larad{v_{z}}$ at the different height.
We also defined the horizontal divergence $D_{\mr{h}}$ and the vertical vorticity $W_{z}$
of the mean field as follows:
\begin{equation}
  D_{\mr{h}}=\frac{\partial }{\partial x}\larad{v_{x}}
  +\frac{\partial }{\partial y}\larad{v_{y}},
  \label{eq:dh}
\end{equation}
\begin{equation}
  W_{z}=\frac{\partial }{\partial x}\larad{v_{y}}
  -\frac{\partial }{\partial y}\larad{v_{x}}.
  \label{eq:wz}
\end{equation}
Figures \ref{fig:div} and \ref{fig:rot} show $D_{\mr{h}}$ and $W_{z}$ at different heights.
We calculated the correlation coefficients (CCs) of $\larad{B_{\mr{np}}}$
versus $\larad{v_{z}}$, $D_{\mr{h}}$, and $W_{z}$ at each height, respectively.
Figure \ref{fig:cor} shows the CCs as a function of height.
The sign of the CCs corresponds to the sign of $\larad{v_{z}}$, $D_{\mr{h}}$, and $W_{z}$
because $\larad{B_{\mr{np}}}$ is always positive.
$\larad{B_{\mr{np}}}$ had a negative correlation with $\larad{v_{z}}$
over a broad range from $z=-100~\mr{Mm}$ to $-20~\mr{Mm}$ but peaked
around $z=-40~\mr{Mm}$ ($\mr{CC}\sim -0.7$).
This indicates the relationship
between the nonpotential field in the photosphere and the downflow rooted to the deep layer 
in the convection zone.
$\larad{B_{\mr{np}}}$ had a weak or no correlation with $D_{\mr{h}}$ and $W_{z}$.

\begin{figure}
  \begin{center}
    \includegraphics[width=\columnwidth]{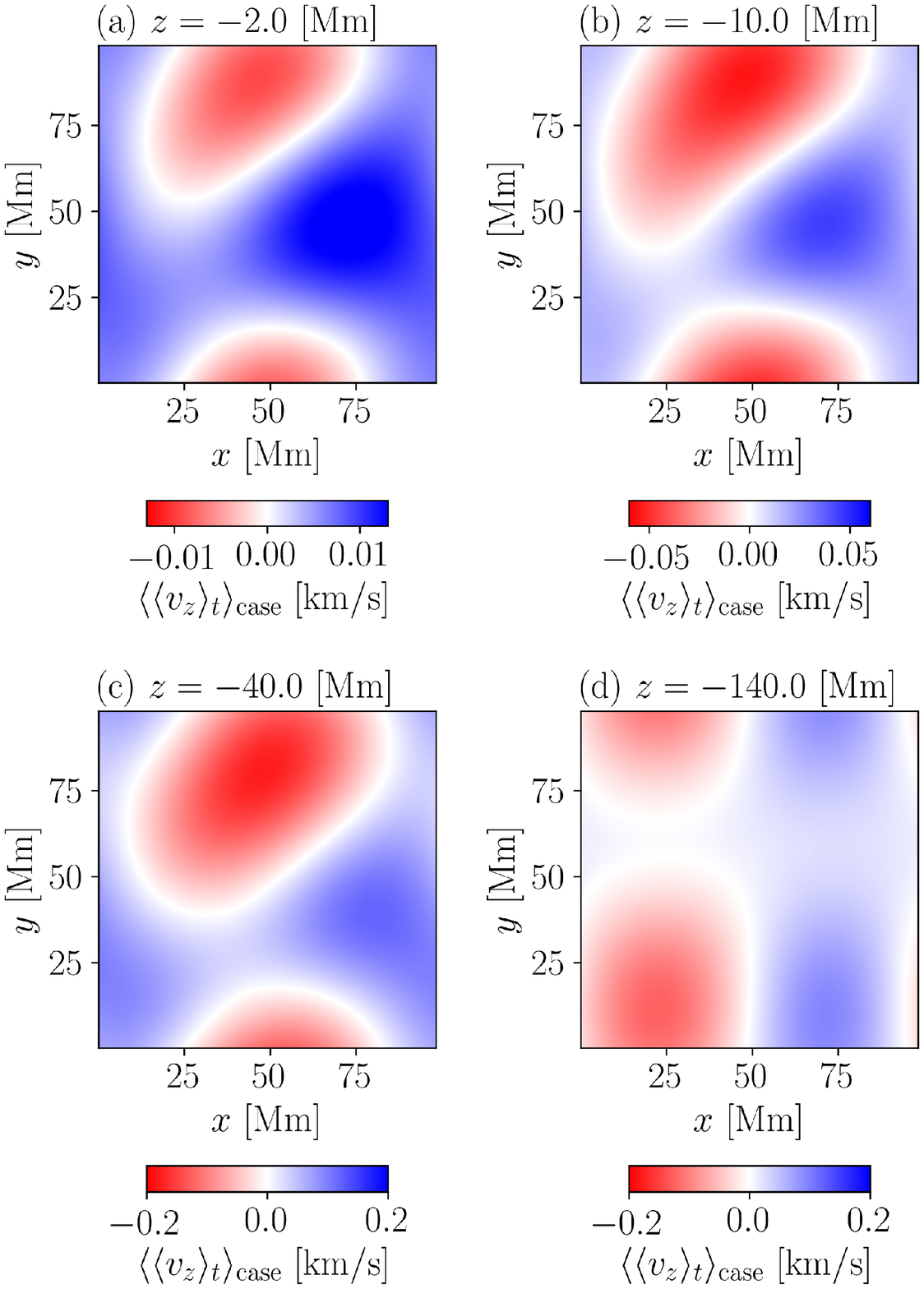}
    \caption{$\larad{v_{z}}$ excluding small-scale structures using a low-pass filter with
      $f_{\mr{cutoff}}=1.44\times 10^{-2}~\mr{Mm^{-1}}$.
      The panels show the filtered $\larad{v_{z}}$ at different heights.}
      \label{fig:vxs_lp}
  \end{center}
\end{figure}

\begin{figure}
  \begin{center}
    \includegraphics[width=\columnwidth]{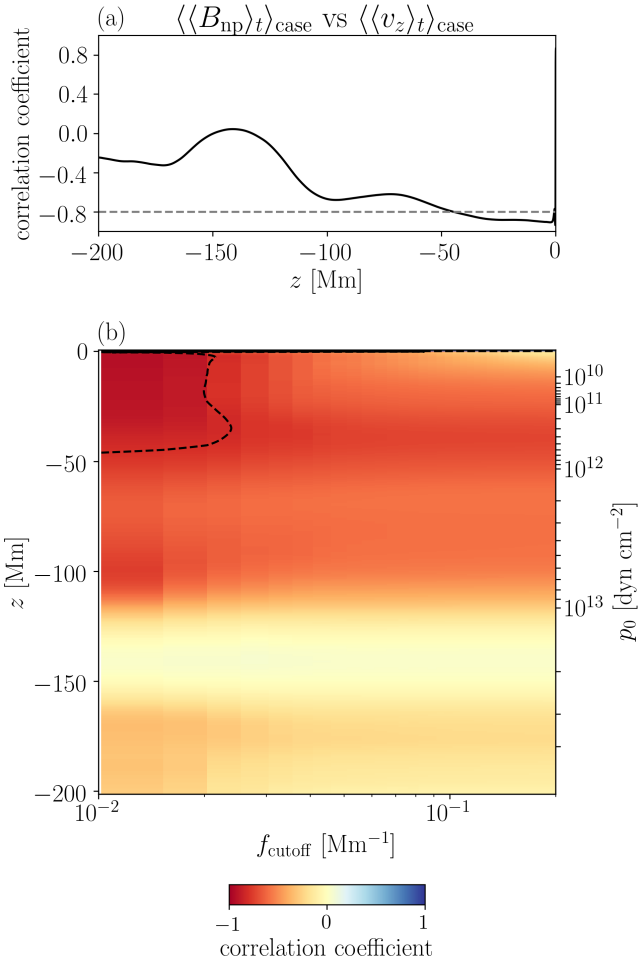}
    \caption{Correlation between filtered $\larad{v_{z}}$ and $\larad{B_{\mr{np}}}$.
      Panel (a) shows the correlation coefficients as a function of height with
      $f_{\mr{cutoff}}=1.44\times 10^{-2}~\mr{Mm^{-1}}$.
      Panel (b) shows the correlation coefficients as a function of height and $f_{\mr{cutoff}}$.
      The dashed lines in panels (a) and (b) denote $\mr{CC}=-0.8$.}
      \label{fig:cor_vx_lp}
  \end{center}
\end{figure}

It is likely that small-scale fluctuations reduce CCs to smaller values.
We excluded the small-scale structures from $\larad{v_{z}}$, $D_{\mr{h}}$, and $W_{z}$
using a low-pass filter.
The low-pass filter was applied in the two-dimensional horizontal plane at each height.
Figure \ref{fig:vxs_lp} shows the filtered $\larad{v_{z}}$ including
$f<f_{\mr{cutoff}}=1.44\times 10^{-2}~\mr{Mm^{-1}}$
where $f(=1/\lambda )$ represents the wavenumber ($\lambda $ represents the wavelength)
and $f_{\mr{cutoff}}$ is the cutoff wavenumber.
The distributions at $z=-2.0~\mr{Mm}$, $-10~\mr{Mm}$, and $-40~\mr{Mm}$
(panels (a), (b), and (c)) are similar to each other.
Figure \ref{fig:cor_vx_lp} (a) shows the correlation between $\larad{B_{\mr{np}}}$ and the filtered $\larad{v_{z}}$
as a function of height.
It shows a strong correlation ($|\mr{CC}|>0.8$) in $x>-48~\mr{Mm}$.
Figure \ref{fig:cor_vx_lp} (b) shows the CC as a function of height and cutoff wavenumber
of the low-pass filter.
$\mr{|CC|>0.8}$ is found in the parameter space of $-48~\mr{Mm}<x<0$
and $f_{\mr{cutoff}}<0.02~\mr{Mm^{-1}}$
(inside the dashed line in Fig. \ref{fig:cor_vx_lp} (b))
corresponding to the region and the spatial scale of the downflow plume $\sim 50~\mr{Mm}$.

\begin{figure}
  \begin{center}
    \includegraphics[width=\columnwidth]{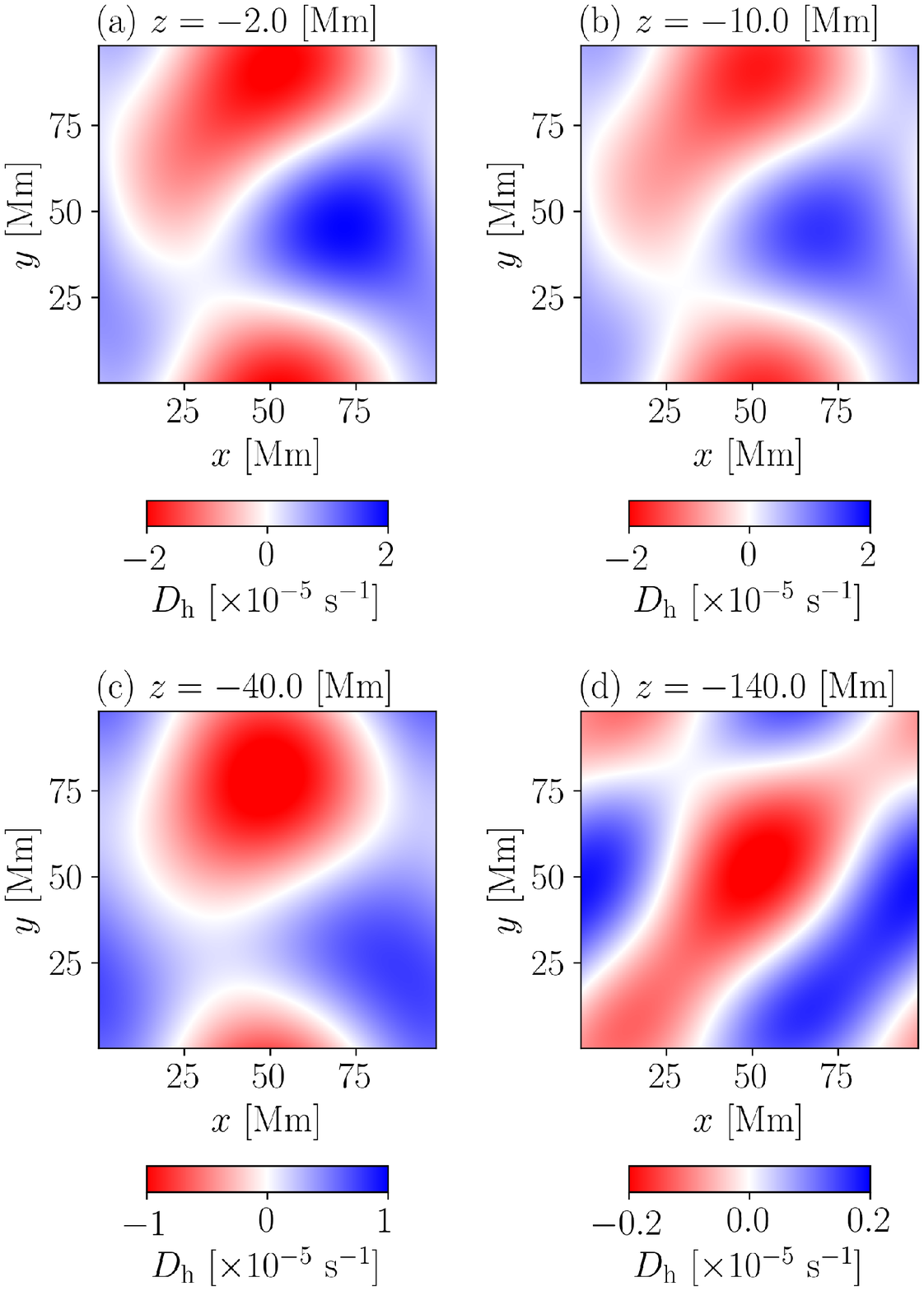}
    \caption{Same as Fig. \ref{fig:vxs_lp} but for $D_{\mr{h}}$.}
    \label{fig:div_lp}
  \end{center}
\end{figure}

\begin{figure}
  \begin{center}
    \includegraphics[width=\columnwidth]{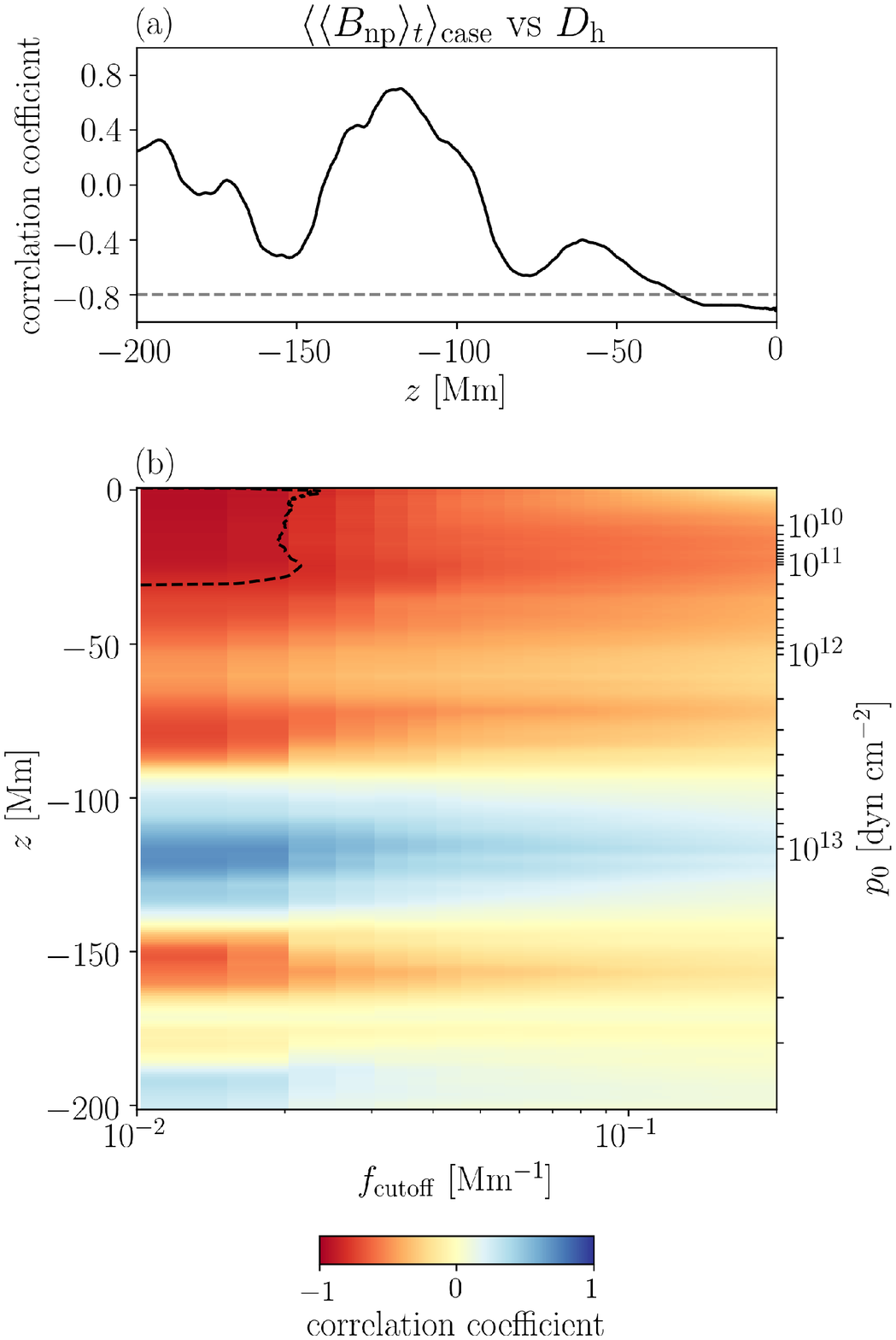}
    \caption{Same as Fig. \ref{fig:cor_vx_lp} but for $D_{\mr{h}}$.}
      \label{fig:cor_div_lp}
  \end{center}
\end{figure}

Figure \ref{fig:div_lp} shows the filtered $D_{\mr{h}}$ including
$f<f_{\mr{cutoff}}=1.44\times 10^{-2}~\mr{Mm^{-1}}$.
The filtered $D_{\mr{h}}$ shared the similar distribution
at $z=-2.0~\mr{Mm}$, $-10~\mr{Mm}$, and $-40~\mr{Mm}$ (Fig. \ref{fig:div_lp} (a), (b), and (c)).
They were also similar to the filtered $\larad{v_{z}}$
at $z=-2.0~\mr{Mm}$, $-10~\mr{Mm}$, and $-40~\mr{Mm}$ (Fig. \ref{fig:vxs_lp} (a), (b), and (c)).
Figure \ref{fig:cor_div_lp} shows the correlation between $\larad{B_{\mr{np}}}$
and the filtered $D_{\mr{h}}$.
$\mr{CC}<-0.8$, which indicates a strong correlation between the nonpotential field
and horizontal converging flows, was found at
$z>-30~\mr{Mm}$ and $f_{\mr{cutoff}}<0.02~\mr{Mm^{-1}}$.

In the deep convection zone, the anelastic approximation $\nabla \cdot (\rho _{0}\vctr{v})=0$
is valid. The relationship between the downflow plume and the horizontal converging flows
is formulated as follows \citep{2014ApJ...793...24L}:
\begin{eqnarray}
  0 &=& \nabla \cdot (\rho _{0}\vctr{v}) \nonumber, \\
  ~ &=& \nabla _{\mr{h}}\cdot (\rho _{0}\vctr{v}_{\mr{h}})+\partial _{z} (\rho _{0}v_{z}) \nonumber, \\
  ~ &=& \rho _{0}\nabla _{\mr{h}}\cdot \vctr{v}_{\mr{h}} +v_{z}\partial _{z}\rho _{0} + \rho _{0}\partial _{z}v_{z}\nonumber, \\
  \Leftrightarrow \nabla _{\mr{h}}\cdot \vctr{v}_{\mr{h}} &=& \frac{v_{z}}{H_{\rho }} - \partial _{z}v_{z}.
\end{eqnarray}
where $\vctr{v}_{\mr{h}}$ represents the horizontal velocity and
$H_{\rho }=-( \partial \ln \rho _{0}/\partial z)^{-1}$ is the scale height.
We assume $v_{z}/H_{\rho } \gg \partial _{z}v_{z}$ for the downflow whose spatial scale is much larger
than the scale height. Thus, we obtain the following equation:
\begin{equation}
  \nabla _{\mr{h}}\cdot \vctr{v}_{\mr{h}}=\frac{v_{z}}{H_{\rho }}. \label{eq:mass}
\end{equation}
Equation (\ref{eq:mass}) indicates that large-scale downflow (negative $v_{z}$) is followed
by the horizontal converging flows.
$D_{\mr{h}}$ and $\larad{B_{\mr{np}}}$ are strongly correlated
when $v_{z}$ is strongly correlated with $\larad{B_{\mr{np}}}$.

\begin{figure}
  \begin{center}
    \includegraphics[width=\columnwidth]{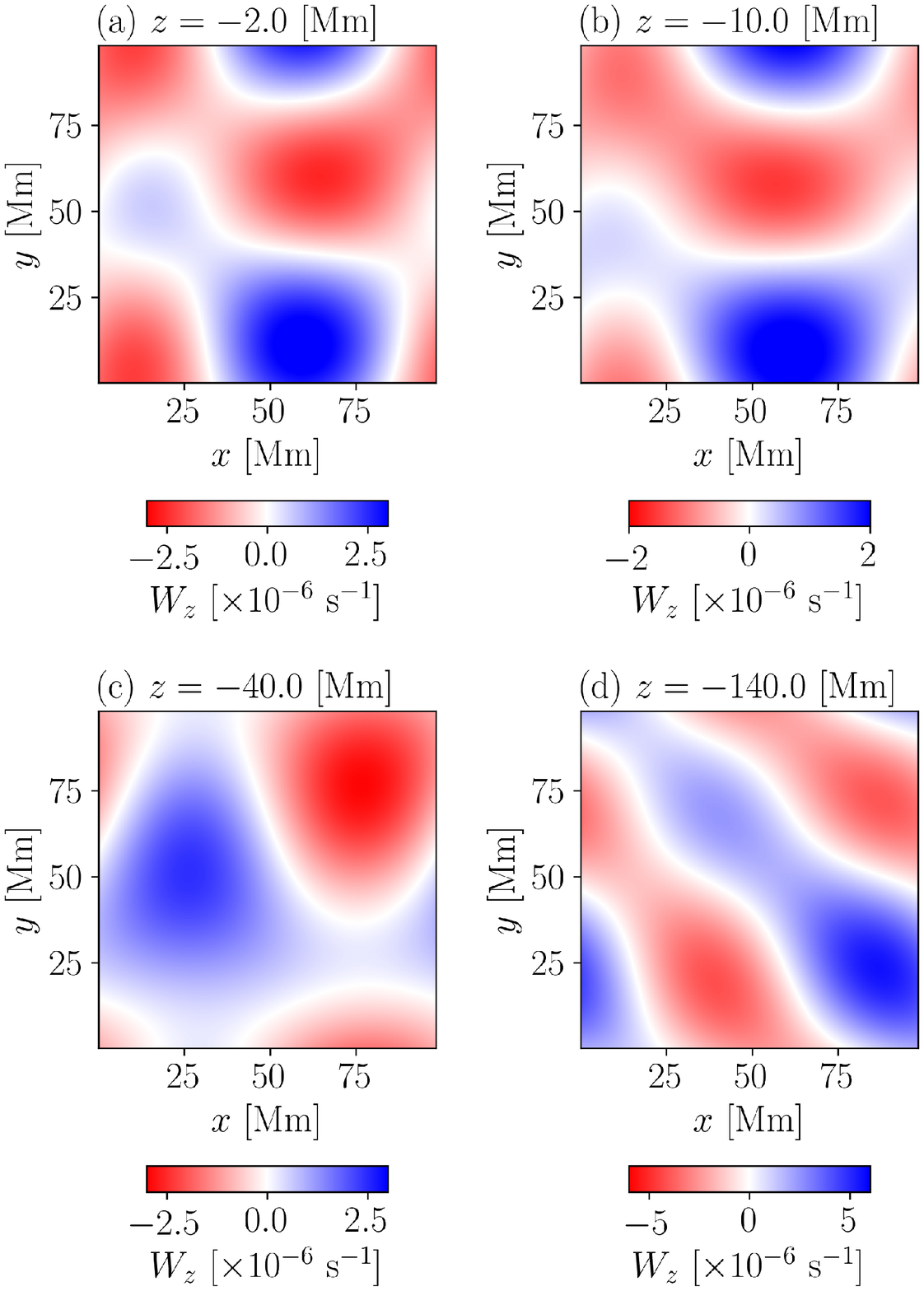}
    \caption{Same as Fig. \ref{fig:vxs_lp} but for $W_{z}$.}
      \label{fig:rot_lp}
  \end{center}
\end{figure}

\begin{figure}
  \begin{center}
    \includegraphics[width=\columnwidth]{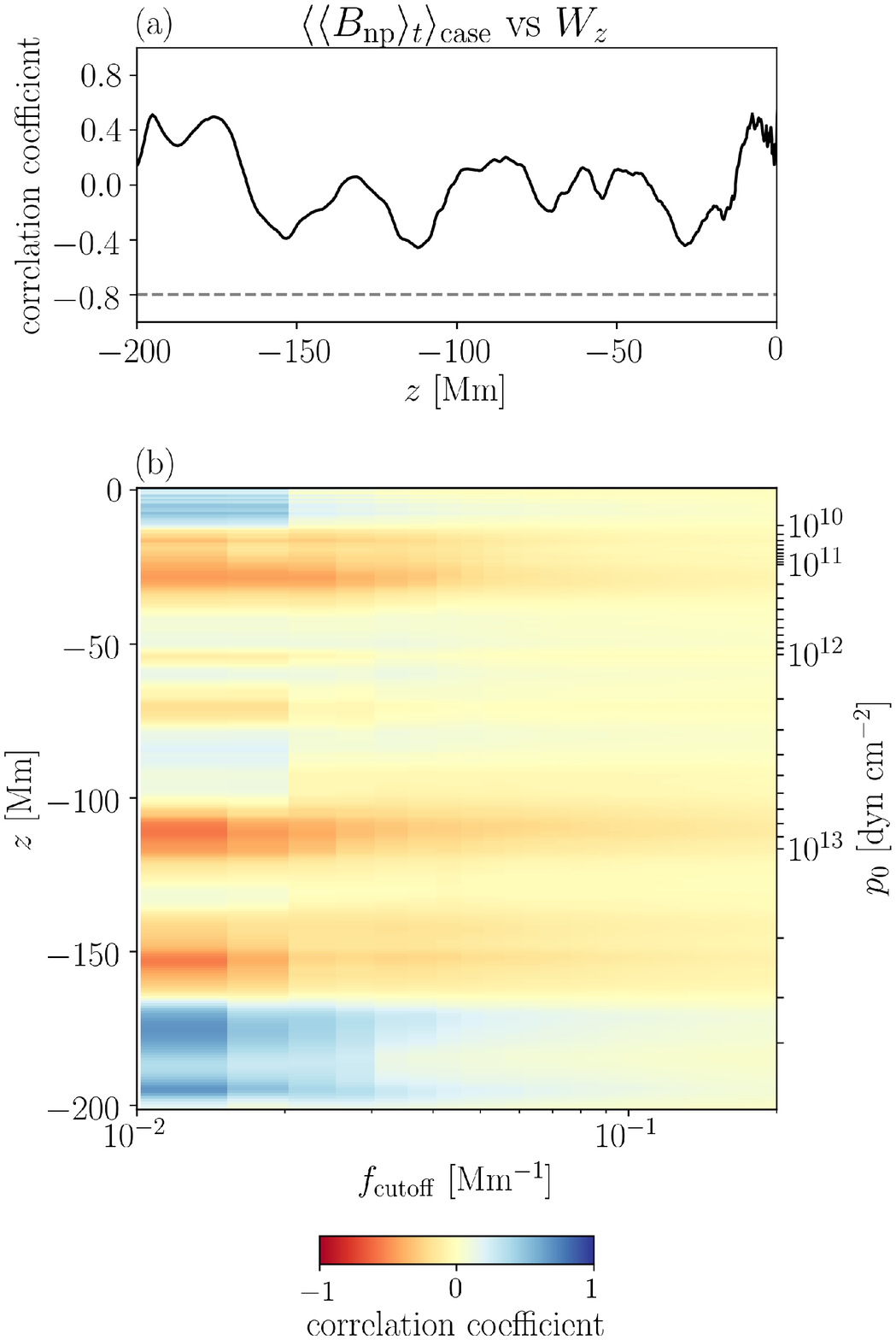}
    \caption{Same as Fig. \ref{fig:cor_vx_lp} but for $W_{z}$}
    \label{fig:cor_rot_lp}
  \end{center}
\end{figure}

Figure \ref{fig:rot_lp} shows the filtered $W_{z}$ including
$f<f_{\mr{cutoff}}=1.44\times 10^{-2}~\mr{Mm^{-1}}$.
The distributions at $z=-2~\mr{Mm}$ and $z=-10~\mr{Mm}$ are similar,
but the distributions at $z=-40~\mr{Mm}$ and $z=-140~\mr{Mm}$
are different from the two shallow distributions.
The distributions are also different from the filtered $\larad{v_{z}}$ and $D_{\mr{h}}$ distributions
in Figs. \ref{fig:vxs_lp} and \ref{fig:div_lp}.
Figure \ref{fig:cor_rot_lp} shows the correlation between $\larad{B_{\mr{np}}}$ 
and filtered $W_{z}$.
The correlation coefficients do not exceed 0.8 at any height or $f_{\mr{cutoff}}$.
This result suggests that the vertical vortices of the convective flows were not the origin of the
$\larad{B_{\mr{np}}}$ distribution.

\begin{figure}
  \begin{center}
    \includegraphics[width=\columnwidth]{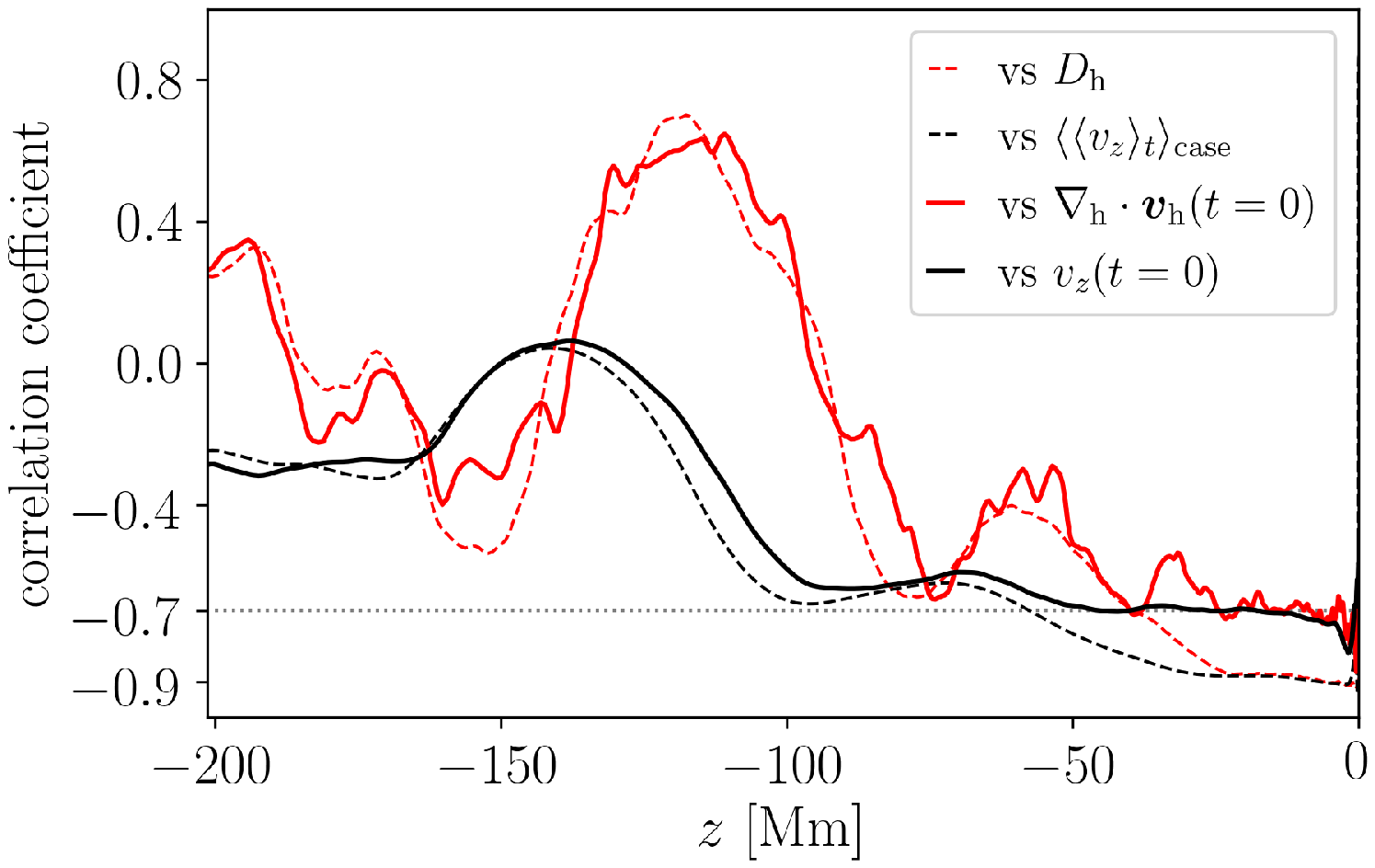}
    \caption{Correlation coefficients as a function of height.
      Black solid line: $v_{z}(t=0)$ vs $\larad{B_{\mr{np}}}$,
      red solid line: lowpass-filtered $\nabla _{\mr{h}} \cdot \vctr{v}_{\mr{h}} (t=0)$ vs $\larad{B_{\mr{np}}}$,
      black dashed line: $\larad{v_{z}}$ vs $\larad{B_{\mr{np}}}$ (same as Fig. \ref{fig:cor_vx_lp} (a)),
      red dashed line: $D_{\mr{h}}$ vs $\larad{B_{\mr{np}}}$ (same as Fig. \ref{fig:cor_div_lp} (a)).}
      \label{fig:cor_t0}
  \end{center}
\end{figure}

We also compared $\larad{B_{\mr{np}}}$ with the convective velocity field
at the initial state $t=0$.
The convective velocity field at $t=0$, common for all cases,
was not affected by the magnetic field.
Figure \ref{fig:cor_t0} shows the CCs of the filtered distributions
of $v_{z}$ and $\nabla _{\mr{h}} \cdot \vctr{v}_{\mr{h}}$
($f_{\mr{cutoff}}=1.44\times 10^{-2}~\mr{Mm^{-1}}$) against $\larad{B_{\mr{np}}}$.
CCs in $x>-50~\mr{Mm}$ are approximately $-0.7$.
This result suggests that the place creating the photospheric nonpotential fields
depends on the distribution of the downflow plume extending to the convection zone
within the depth of $-50~\mr{Mm}$.
The results also suggest that the photospheric nonpotential field distribution can be predicted
by detecting the downflow plume or horizontal converging flows
even before the flux appears in the photosphere.

\section{Summary and Discussion} \label{sec:dis}
We carried out 93 radiative MHD simulations reproducing the transport of the magnetic
flux tubes in the solar convection zone to the photosphere.
Various types of magnetic distributions including $\delta $-type and $\beta $-type were
reproduced in the photosphere only by the difference in the convective flows
surrounding the flux tubes.
In this study, we adopted the convective velocity field including a persistent large-scale
downflow plume as the initial state.
In the cases where the flux tubes were trapped by the downflow plume,
the $\delta $-type magnetic distribution
with the strong nonpotential field was created in the photosphere
due to the collision of positive and negative fluxes.
The previous studies \citep{2009ApJ...697.1529F,2015A&A...582A..76S,2019ApJ...886L..21T}
found the unipolar rotation of the sunspots transporting the magnetic helicity
from the convection zone to the corona.
We newly found the bipolar rotation of the $\delta$-spots
whose direction is opposite to the unipolar rotation.
The bipolar rotation was driven by the conversion of twist to writhe.
In our parameter survey, the $\delta $-type magnetic distribution was more frequently observed than
the $\beta $-type distribution.
There were also cases without flux emergence when the initial position of the flux tubes was
just below the downflow plumes, and most of the flux tube body was filled with downflow.

Using the results of the statistical analysis, we derived the distribution 
of the nonpotential field $\larad{B_{\mr{np}}}$ in the photosphere.
The comparison with the mean velocity field in the convection zone revealed that $\larad{B_{\mr{np}}}$
was strongly correlated with the distribution of the downflow plumes in the convection zone.
The distribution of the horizontal converging flows
also showed a strong correlation with $\larad{B_{\mr{np}}}$.
The results support the scenario that the nonpotential field is formed
by the collision of opposite-polarity magnetic fluxes trapped by the downflow plume
\citep{2019ApJ...886L..21T,2020MNRAS.498.2925H}.
The horizontal converging flow gathers the emerging flux tubes into the downflow region.
Hence, the destination of the emerging flux tube was most likely above the downflow plumes.

In the analyses, we regarded the $\tau =1$ surface as a flat surface.
We evaluated the impact of this assumption on our results.
In case 001, for example,
the average $\tau =1$ surface was $67~\mathrm{km}$ at $t=50~\mathrm{h}$.
The grid point closest to the average $\tau =1$ surface was at $z=52~\mathrm{km}$.
The difference of the magnetic flux was
$|\Phi ^{\tau }-\Phi ^{z}|/\Phi ^{z}=0.06$,
where the superscripts $\tau$ and $z$ represent the values
at the average $\tau =1$ surface and at $z=52~\mathrm{km}$, respectively.
The mean squared errors of
$B_{x}^{\tau}$ vs $B_{x}^{z}$,
$B_{y}^{\tau}$ vs $B_{y}^{z}$, and
$B_{z}^{\tau}$ vs $B_{z}^{z}$, 
were $50.0~\mathrm{G}$, $63.3~\mathrm{G}$, and $50.2~\mathrm{G}$, respectively.
The mean squared error of
$B_{\mathrm{np}}^{\tau }$ vs $B_{\mathrm{np}}^{z}$
was $99.4~\mathrm{G}$.
The correlation coefficients of
$B_{x}^{z}$ vs $B_{x}^{\tau }$,
$B_{y}^{z}$ vs $B_{y}^{\tau }$, and 
$B_{z}^{z}$ vs $B_{z}^{\tau }$
were $0.98$, $0.91$, and $0.95$, respectively.
The correlation coefficient of
$\lara{B_{\mathrm{np}}}_{t}^{z}$ vs $\lara{B_{\mathrm{np}}}_{t}^{\tau}$ was 0.9.
In the other cases as well as case 001,
the difference of the magnetic flux was a few percent, and the correlation coefficient
between $\lara{B_{\mathrm{np}}}_{t}^{z}$ vs $\lara{B_{\mathrm{np}}}_{t}^{\tau}$
was over $0.9$. Therefore, we conclude that
the magnetic fluxes and the correlation
between $\larad{B_{\mathrm{np}}}$ and $\larad{v_{z}}$ are not significantly changed
if we compute those values using the magnetic field data at a certain height
close to the solar surface.

The correlation between $\larad{B_{\mr{np}}}$ and
the distribution of the downflow plume before flux emergence ($|\mr{CC}|\sim 0.7$) suggests that high free energy regions in the photosphere can be predicted
before flux emergence by detecting the downflow plume in the convection zone.
In our simulation, $t=0$ was $20$--$30$ h before flux emergence;
i.e., the leading time was almost one day.
The leading time depends on the persistence of the downflow plume.
Previous studies that used helioseismic techniques
detected the signals of the emerging fluxes in the subsurface layer
before the flux appeared in the photosphere \citep{2011Sci...333..993I,2013ApJ...762..131B,2013ApJ...770L..11T}.
The signals were not detected in the quiet Sun without flux emergence.
\citet{2011Sci...333..993I} reported that the signals of the emerging fluxes were
mainly contributed by the acoustic waves at the depth of $57$--$60~\mathrm{Mm}$,
and they were concentrated in the area with a size of $30$--$50~\mathrm{Mm}$ in the horizontal direction.
Our simulation results were quantitatively consistent with the observational results of previous studies.

The detection of the pre-emerging regions is still challenging.
We have to select local candidate areas to apply the helioseismic analysis
in the solar hemisphere;
however, there are few clues to determine them before the flux emergence.
Our findings are helpful to limit the regions
where helioseismic analysis should be applied
and contribute to increase the efficiency for successful detection
of pre-emerging regions in the solar hemisphere.
Helioseismology can be used to detect the horizontal flows rather than the vertical flows.
Hence, the correlation between the horizontal converging flows and $\larad{B_{\mr{np}}}$
can improve the practical detection.
Even if we cannot detect the flux emergence beforehand,
we can still predict that the emerging flux most likely develops into a $\delta$ -spot.

In actual observations, $\beta $-spots are more frequently detected, whereas
in our parameter survey, $\delta $-spots were more frequently observed.
Two factors determine where the $\delta $-spots favorably form.
One is the downflow plume in the convection zone; the other factor is the birthplace of the flux tubes.
In this study, the initial positions of the flux tubes were assumed to be uniformly distributed,
and thus, there is the possibility that the flux tubes are located
in the regions where they do not actually exist.
Our results imply that there should be favorable locations of the birthplace of the flux tubes.
The theoretical findings and previous numerical simulations suggested that the toroidal magnetic field of the flux tubes
can be created at the overshoot region in the tachocline \citep{2021LRSP...18....5F}.
The exact region where the flux tubes are generated is still under debate.
Further investigation using dynamo simulation is required to constrain the birthplace of the flux tubes.
The periodic boundary in our simulations can also be the reason why the $\delta$-spots were more frequently created.
Due to the periodic boundary, magnetic fluxes that left from one side of the simulation domain reentered from the opposite side, leading to the higher frequency of the flux collision. This process occurred in \citet{2019ApJ...886L..21T}, but it was not always the case in this study.

As shown in Fig. \ref{fig:flux}, our results covered relatively larger amount of magnetic fluxes.
The flux emergence rates in our results were several times larger than the observational values,
resulting in larger scaling exponents than the previous observational results of
\citet{2011PASJ...63.1047O}.
Most of the simulated results including those of the previous studies showed
larger emergence rates because the simulated turbulent velocity was faster
than the realistic value, and the initial flux tube model was simple.
We assumed the straight flux tube with uniform field strength along the axis in the initial state,
but the actual flux tube created in the convection zone should have
the nonuniform field strength and the curved geometry.
To include the realistic flux tube model,
the numerical setting adopted in \citet{2017ApJ...846..149C},
where the initial flux tube model referred a result of a dynamo simulation, should be used.
The other possibility is that the observational values had too large dispersion
to obtain the robust scaling exponent.
Our simulated results, where the initial turbulent velocity fields were common,
showed relatively smaller dispersion.
The observational values in Fig. \ref{fig:flux} include the results from different active regions
in different places and times; thus, turbulence conditions can be different
in each case. The method of data analysis can also affect the results, e.g.,
\citet{2011PASJ...63.1047O} used the data of
the maximum $\Phi $ versus the temporally averaged $d\Phi/dt$ for the fitting,
while we used the data of the maximum $\Phi $ versus the maximum $d\Phi/dt$.    
Further statistical studies are required to unravel the impacts of turbulence and magnetic geometry
on the relationship between $\Phi $ and $d\Phi /dt$ from both observational and numerical perspectives.

\section*{Acknowledgements}
We thank the referee for the constructive comments.
This work was supported by MEXT as ``Program for Promoting Researches on the Supercomputer Fugaku''
  (Toward a unified view of the universe: from large scale structure to planets,
  Elucidation of solar and planetary dynamics and evolution, grant no. 20351188),
  JSPS KAKENHI Grant Nos. JP20KK0072 (PI: S. Toriumi), JP21H01124 (PI: T. Yokoyama),
  JP20K14510 (PI: H. Hotta), JP21H04497 (PI: H. Miyahara)
  and JP21H04492 (PI: K. Kusano),
  and the NINS program for cross-disciplinary study (Grant Nos. 01321802 and 01311904)
  on Turbulence, Transport, and Heating Dynamics in Laboratory and Astrophysical Plasmas: “SoLaBo-X.
  TK was supported by the National Center for Atmospheric Research,
  which is a major facility sponsored by the National Science Foundation under Cooperative Agreement No. 1852977.
  A part of this study was carried out using the computational resources of
  the Center for Integrated Data Science, Institute for Space-Earth Environmental Research, Nagoya University. We would like to thank Editage (www.editage.com) for English language editing.

\section*{Data Availability}
Data available on request.











\bsp	
\label{lastpage}
\end{document}